\documentclass[12pt, draftclsnofoot, onecolumn]{IEEEtran}
\usepackage{setspace}
\usepackage{epsfig}
\usepackage{amsmath}
\usepackage{amsfonts}
\usepackage{multicol,flushend,color}
\usepackage{float}

\begin{document}
\title{Optimal Power Allocation by Imperfect Hardware Analysis in Untrusted Relaying Networks}
\author{Ali Kuhestani, \IEEEmembership{Student Member,{\hspace{.7mm}}IEEE}, Abbas Mohammadi, \IEEEmembership{Senior Member,{\hspace{.7mm}}IEEE}, Kai-Kit Wong, \IEEEmembership{Fellow,~IEEE}, Phee Lep Yeoh, \IEEEmembership{Member,~IEEE}, \\Majid Moradikia, and Muhammad R. A. Khandaker,~\IEEEmembership{Member,~IEEE}\\
\thanks{A. Kuhestani and A. Mohammadi are with the Electrical Engineering Department, Amirkabir University of Technology, Tehran, Iran.}
\thanks{K.-K Wong and M. R. A. Khandaker are with the Department of Electronic and Electrical
Engineering, University College London, London WC1E 6BT, U.K.}
\thanks{P. L. Yeoh is with the School of Electrical and Information Engineering,
The University of Sydney, NSW, Australia.}
\thanks{M. Moradikia is with the Electrical Engineering Department, Shiraz University of Technology, Shiraz, Iran.}
}
\maketitle{}
\markboth{}{Ali Abbas}
\begin{abstract}
By taking a variety of realistic hardware imperfections into consideration, we propose an optimal power allocation (OPA) strategy to maximize the instantaneous secrecy rate of a cooperative wireless network comprised of a source, a destination
and an untrusted amplify-and-forward (AF) relay. We assume that either the source or the destination is equipped with a
large-scale multiple antennas (LSMA) system, while the rest are equipped with a single antenna. To prevent the untrusted relay
from intercepting the source message, the destination sends an intended jamming noise to the relay, which is referred to as
destination-based cooperative jamming (DBCJ). Given this system model, novel closed-form expressions are presented in the high signal-to-noise ratio (SNR) regime for the ergodic secrecy rate (ESR) and the secrecy outage probability (SOP). We further improve the secrecy performance of the system by optimizing the associated hardware design. The results reveal that by beneficially distributing the tolerable hardware imperfections across the transmission and reception radio-frequency (RF) front ends of each node, the system's secrecy rate may be improved. The engineering insight is that equally sharing the total imperfections at the relay between the transmitter and the receiver provides the
best secrecy performance. Numerical results illustrate that the proposed OPA together with the most appropriate hardware design significantly increases the secrecy rate.
\end{abstract}
\section{Introduction}
\IEEEPARstart{S}{ecurity} in wireless communication networks is conventionally implemented above the physical layer using key based cryptography {\cite {mukh}}. To complement these highly complex schemes, wireless transmitters can also be
validated at the physical layer by exploiting the dynamic characteristics of the associated communication links \cite {review3}, {\cite {review1}}. Physical layer security (PLS) is a promising paradigm for safeguarding fifth-generation (5G) wireless communication networks without incurring additional security overhead {\cite {review1}}.

Massive multiple-input multiple-output (MIMO) systems as a key enabling technology of 5G wireless communication networks provide significant performance gains in terms of spectral efficiency and energy efficiency {\cite {1}}, {\cite {1.5}}. This new technology employs coherent processing across arrays of hundreds or even thousands of base station (BS) antennas and supports tens or hundreds of mobile terminals {\cite {1}}--{\cite {2}}. As an additional advantage, massive MIMO is inherently more secure than traditional MIMO systems, as the large-scale antenna array exploited at the transmitter can precisely aim a narrow and directional information beam towards the intended receiver, such that the received signal-to-noise ratio (SNR) is several orders of magnitude higher than that at any incoherent passive eavesdropper {\cite {3}}. However, these security benefits are severely hampered in cooperative networks where the intended receivers may also be potential eavesdroppers  {\cite {He}}--{\cite {Khandaker2}}.

In the context of PLS, cooperative jamming which involves the transmission of additional jamming signals to degrade the received SNR at the potential eavesdropper can be applied by source {\cite {Khandaker1}}, {\cite {Khandaker2}}, the intended receiver node {\cite {He}} or a set of nodes, i.e., source and destination or source and relay to beamform the jamming noise orthogonal to the spatial dimension of the desired signal {\cite {Petro}}, {\cite {Petro2}}. Recently, several works have considered the more interesting scenario of untrusted relaying {\cite {Mukherjee}--\cite {sec-aware}} where the cooperative jamming is performed by the intended receiver, which is referred to as {\it destination-based cooperative jamming} (DBCJ).

In real life, an {\it untrusted}, i.e., honest-but-curious, relay may collaborate to provide a reliable communication. Several practical scenarios may include untrusted relay nodes, e.g., in ultra-dense heterogenous wireless networks where low-cost intermediate nodes may be used to assist the source-destination transmission. In these networks, it is important to protect the confidentiality of information from the untrustworthy relay, while concurrently relying on it to increase the reliability of communication. Thanks to the DBCJ strategy {\cite {He}}, positive secrecy rate can still be attained in untrusted relay networks. In recent years, several works have focused on the performance analysis {\cite {Mukherjee}--\cite {Asma}}, power allocation {\cite {Kuhestani2}}--{\cite {error}} and security enhancement {\cite {Kuhestani3}}, {\cite {sec-aware}} of untrusted relaying networks. To be specific, the authors in {\cite {Kuhestani2}} proposed an optimal power allocation (OPA) strategy to maximize the instantaneous secrecy rate of one-way relaying network while two-way relaying scenario was considered in {\cite {Kuhestani3}}. By exploiting the direct link, a source-based artificial noise injection scheme was proposed in {\cite {Lv}} to hinder the untrusted relay from intercepting the confidential message. A power allocation strategy was also proposed in {\cite {Lv}} to optimally determine the information and jamming signal powers transmitted by the source. In {\cite {error}}, the OPA problem with imperfect channel state information was investigated. Notably, all the aforementioned works considered perfect hardware in the communication network.

In practice, hardware equipments experience detrimental impacts of phase noise, I/Q imbalance, amplifier non-linearities, quantization errors, converters, mixers, filters and oscillators {\cite {Schenk}}, {\cite {Studer}}. Each of the imperfections distorts the signals
in its own way. While hardware imperfections are unavoidable, the severity of the imperfections depends on the quality of the hardware used in the radio-frequency (RF) transceivers. The non-ideal behavior of each component can be modeled in detail for the purpose of designing compensation algorithms, but even after compensation there remain residual transceiver imperfections {\cite {Studer}}. {\it This problem is more challenging especially in high rate systems such as LTE-Advanced and 5G networks exploiting inexpensive equipments} {\cite {Schenk}}. Although most contributions in security based wireless networks have assumed perfect transceiver hardware {\cite {He}}--{\cite {sec-aware}}, or only investigated the impact of particular imperfections such as I/Q imbalance {\cite {karas}} or phase noise {\cite {Zhu1}} in the presence of an external eavesdropper, this paper goes beyond these investigations by considering {\it residual hardware imperfections} in physical layer security design {\cite {Schenk}}--{\cite {Bjor}}.

In this paper, we take into account the OPA in a two-hop amplify-and-forward (AF) untrusted relay network where all the nodes suffer from hardware imperfections and either the source or the destination is equipped with large-scale multiple antennas (LSMA) {\cite {Zhu3}}, {\cite {Chen}} while the other nodes are equipped with a single antenna. The DBCJ protocol is operated in the first phase and then the destination perfectly removes the jamming signal via self interference
cancelation in the second phase. For this system model, the main contributions of the paper are summarized as follows:
\begin{itemize}
    \item Inspired by {\cite {Schenk}}--{\cite {Bjor}}, we first present the generalized system model for transceiver hardware imperfections in our secure transmission network. Based on this, we calculate the received instantaneous signal-to-noise-plus-distortion-ratio (SNDR) at the relay and destination.
      \item  We formulate the OPA between the source and destination that maximizes the instantaneous secrecy  rate of untrusted relaying. Accordingly, novel closed-form solutions are derived for the exact OPA. In addition, new simple solutions are derived for the OPA in the high SNR regime.
      \item According to our OPA solutions, novel compact expressions are derived for the ergodic secrecy rate (ESR) and secrecy outage probability (SOP) in the high signal-to-noise-ratio (SNR) regime that can be applied to arbitrary channel fading distributions. To gain further insights, new closed-form expressions are presented over Rayleigh fading channels. The asymptotic results highlight the presence of a secrecy rate ceiling which is basically different from the prefect hardware case. We highlight that this ceiling phenomenon is independent of the fading characteristic of the two hops.
  \item We provide new insights for hardware design in DBCJ-based secure communications. To this end, under the cost constraint of transceiver hardware at each node, we formulate the hardware design problem for the aforementioned network to maximize the secrecy rate. The results reveal that the secrecy rate can be improved by optimally distributing the level of hardware imperfections between the transmit and receive RF front ends of each node.
\end{itemize}

{\it Notation}: We use bold lower case letters to denote vectors. ${\bf I}_N$ and ${\bf 0}_{N\times 1}$ denote the Identity matrix and the zeros matrix, respectively. $\|.\|$, $(.)^H$ and $(.)^T$ denote the Euclidean norm, conjugate transpose and transpose operators, respectively;
$\mathbb{E}_x\{\cdot\}$ stands for the expectation over the random variable (r.v.) x; $\Pr(\cdot)$ denotes the probability; $f_X(\cdot)$ and $F_X(\cdot)$ denote the probability density function (pdf) and cumulative distribution function (cdf) of the r.v. $X$, respectively; the $\mathcal{CN} (\mu, \sigma^2)$ denotes a circularly symmetric complex Gaussian
RV with mean $\mu$ and variance $\sigma^2$; $\mathrm{diag}({\bf A})$ stands for the main diagonal elements of matrix A; $\mathrm{Ei}(x)$ is the exponential integral  {\cite[Eq. (8.211)] {table}}. $[\cdot]^+=\max\{0,x\}$ and $\max$ stands for the maximum value.
\begin{figure}[t]
  \begin{center}
    \includegraphics[width=6in,height=4in]{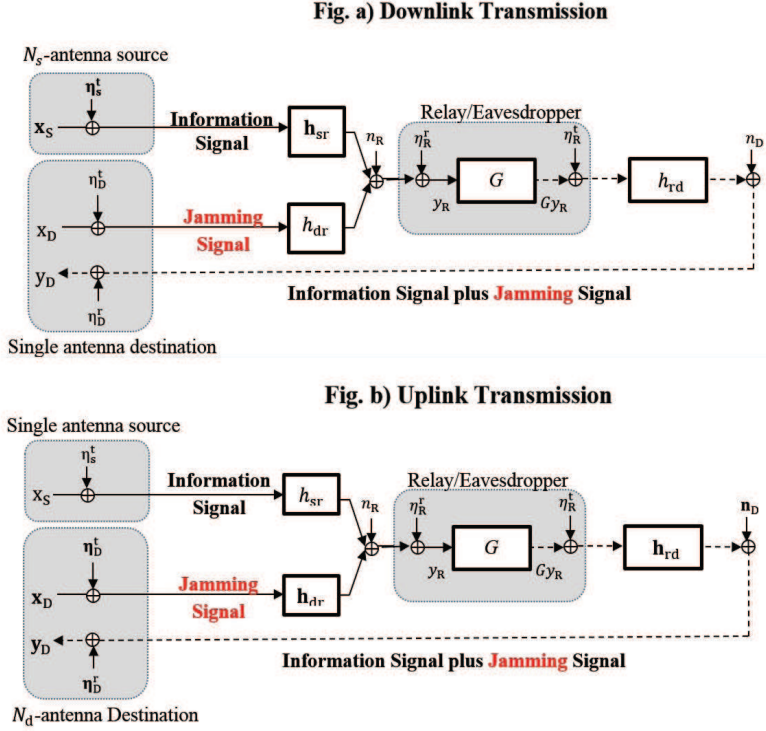} 
    \vspace{-3mm}\caption{Secure transmission under the presence of transceiver imperfections. The first and the second figures show the downlink and the uplink transmissions, respectively. The relay acts as both helper and eavesdropper. The solid lines represent the first phase of transmission while the dashed line represents the second phase of transmission.}\vspace{-5mm}
  \label{system_model}\end{center}
\end{figure}
\section{Signal and System Representation} \label{Sec Model}
\subsection{System Model}

As shown in Fig. \ref{system_model}, the system model under consideration is a wireless network with one source ($\mathrm{S}$), one destination ($\mathrm{D}$) and one untrusted AF relay ($\mathrm{R}$). While $\mathrm{R}$ is equipped with one antenna, $\mathrm{S}$ or $\mathrm{D}$ is equipped with LSMA denoted by $N_\mathrm{s}$ or $N_\mathrm{d}$, respectively {\cite {Chen2}}, {\cite {wang_hanzo}}. This corresponds to the downlink (DL) and uplink (UL) scenarios in a cellular system where the base station is equipped with a LSMA and the mobile user and relay are equipped with a single antenna {\cite {Chen2}}. In the DL scenario, S has many antennas, whereas D has a single antenna. The opposite applies to the UL scenario.

All the nodes operate in a half-duplex mode. Accordingly, D cannot receive the transmitted signal from S while transmitting the jamming
signal and hence, the direct link between S and D is unavailable. We also assume that the channels satisfy the reciprocity theorem {\cite {He}}. In DL transmission, the complex Gaussian channel from $\mathrm{S}$ to $\mathrm{R}$ and $\mathrm{R}$ to $\mathrm{D}$ are denoted by ${\bf h}_{\mathrm{sr}}\sim \mathcal{CN}({\bf 0}_{{N_{\mathrm{s}}}\times 1},\mu_{{\mathrm{sr}}}{\bf {I}}_{N_{\mathrm{s}}})$ and ${h}_{\mathrm{rd}}\sim \mathcal{CN}( 0,\mu_{{\mathrm{rd}}})$, respectively, and for UL transmission, they are denoted by ${h}_{\mathrm{sr}}\sim \mathcal{CN}(0,\mu_{{\mathrm{sr}}})$ and ${\bf h}_{\mathrm{rd}}\sim \mathcal{CN}( {\bf 0}_{{N_{\mathrm{d}}}\times 1},\mu_{{\mathrm{rd}}}{\bf {I}}_{N_{\mathrm{d}}})$, respectively. We consider slow fading such that the channel coefficients vary independently from one frame to another frame and, they do not change within one frame. The additive white noise $n_i$ ($i \in\{\mathrm{R},\mathrm{D}\}$) at each receiver is represented by a zero-mean complex Gaussian variable with variance $N_0$. We define the SNRs per link as $\gamma_{\mathrm{sr}}{\stackrel{\tiny \Delta}{=}}\rho{\Arrowvert {\bf h}_{{\mathrm{sr}}}\Arrowvert^2}$ and $\gamma_{\mathrm{rd}}{\stackrel{\tiny \Delta}{=}}\rho{\Arrowvert {\bf h}_{{\mathrm{rd}}}\Arrowvert^2}$ and hence, the average SNRs per branch is given by $\overline{\gamma}_{\mathrm{sr}}=\rho{\mu_{\mathrm{sr}}}$ and $\overline{\gamma}_{\mathrm{rd}}=\rho{\mu_{\mathrm{rd}}}$, where $\rho=\frac{P}{N_0}$ represents the transmit SNR of the network. The maximum ratio transmission (MRT) beamforming and maximal ratio combining (MRC) processing are applied at the multi-antenna node to improve the overall system performance {\cite {Chen2}}. Let $\nu= \frac{\gamma_\mathrm{sr}}{\gamma_\mathrm{rd}}$ represent the ratio between the source-to-relay and relay-to-destination SNRs. With LSMA, we consider $\nu\gg 1$ in the DL scenario while $\nu \ll 1$ in the UL scenario. These two special cases are taken into account in this paper. As observed in the numerical results, the analysis are satisfactory even for moderate values of $\nu$.

The DBCJ technique is applied to degrade the received signal at the untrusted relay such that it cannot decipher the desired information. The whole transmission is performed based on a time-division multiple-access (TDMA) based protocol such that the message transmission is divided into two phases, i.e. the broadcast phase and the relaying phase. We consider a total transmit power budget for S and D of $P$  with power allocation factor $\lambda\in (0,1)$ such that the transmit powers at S and D are $\lambda P$ and $(1-\lambda)P$, respectively {\cite {sun}}, {\cite {Kuhestani3}}. As such, during the first phase, while S transmits the intended signal with power $\lambda P$, concurrently D jams with a Gaussian noise to confuse the untrusted relay with power $(1-\lambda)P$. For simplicity, the transmit power at R is set to $P$ and accordingly, in the second phase of transmission, R simply broadcasts the amplified version of the received signal with power of $P$.

In order to consider the residual transceiver imperfections (after conventional compensation algorithms have been applied) at node $i$, $i\in \{\mathrm{S}, \mathrm{R}, \mathrm{D}\}$, the generalized system model from {\cite {Schenk}} is taken into account. The imperfection at transmission and reception segments denoted by ${\eta}_{\mathrm{i}}^{t}$ and ${\eta}_{\mathrm{i}}^{r}$ respectively, are introduced as distortion noises. The experimental results in {\cite{Studer}} and many theoretical investigations in {\cite {Studer}}, {\cite {priyanto}} have verified that these distortion noises are well-modeled as Gaussian distributions due to the central limit theorem. A key property is that the variance of distortion noise at an antenna is proportional to the signal power at that antenna {\cite {Studer}}. Accordingly, for the DL scenario, we have {\cite {Studer}}
\begin{eqnarray}\label{inst_sr4}
{\boldsymbol \eta}_{\mathrm{S}}^{t}{\hspace{-3mm}}&\sim&{\hspace{-3mm}}\mathcal{CN}\Big(0,
\frac{\lambda P {k_{{\mathrm{S}}}^{t}}^2}{{\Arrowvert {\bf h}_{{\mathrm{sr}}}\Arrowvert^2}} ~\mathrm{diag}({|{ h}_{{\mathrm{sr}}_1}|^2}~...~{|h_{{\mathrm{sr}}_{N_\mathrm{s}}}|^2})\Big),\nonumber\\
\eta_{\mathrm{D}}^{t}{\hspace{-3mm}}&\sim&{\hspace{-3mm}}\mathcal{CN}\Big(0,
(1-\lambda)P {k_{{\mathrm{D}}}^{t}}^2\Big), \nonumber\\
\eta_{\mathrm{D}}^{r}{\hspace{-3mm}}&\sim&{\hspace{-3mm}}\mathcal{CN}\Big(0,
P {k_\mathrm{D}^r}^2 |h_\mathrm{rd}|^2\Big).
\end{eqnarray}
Furthermore, we have
\begin{eqnarray}\label{inst_sr4_2}
\eta_{\mathrm{S}}^{t}{\hspace{-3mm}}&\sim&{\hspace{-3mm}}\mathcal{CN}\Big(0,
\lambda P {k_{{\mathrm{s}}}^{t}}^2\Big),\nonumber\\
{\boldsymbol \eta}_{\mathrm{D}}^{t}{\hspace{-3mm}}&\sim&{\hspace{-3mm}}\mathcal{CN}\Big(0,
\frac{(1-\lambda)P{k_{{\mathrm{D}}}^{t}}^2}{{\Arrowvert {\bf h}_{{\mathrm{rd}}}\Arrowvert^2}}~ \mathrm{diag} ({|h_{{\mathrm{rd}}_1}|^2}~...~{|h_{{\mathrm{rd}}_{N_\mathrm{d}}}|^2})\Big), \nonumber\\
{\boldsymbol \eta}_{\mathrm{D}}^{r}{\hspace{-3mm}}&\sim&{\hspace{-3mm}}\mathcal{CN}\Big(0,
P {k_\mathrm{D}^r}^2 \mathrm{diag} ({|h_{{\mathrm{rd}}_1}|^2}~...~{|h_{{\mathrm{rd}}_{N_\mathrm{d}}}|^2})\Big),
\end{eqnarray}
for the UL scenario. The imperfections at R of both cases are also given by
\begin{eqnarray}
\eta_{\mathrm{R}}^{t}{\hspace{-3mm}}&\sim&{\hspace{-3mm}} \mathcal{CN} \Big(0, P{k_{{\mathrm{R}}}^{t}}^2\Big),\nonumber\\
\eta_{\mathrm{R}}^{r}{\hspace{-3mm}}&\sim& {\hspace{-3mm}}\mathcal{CN}\Big(0,P{k_{{\mathrm{R}}}^{r}}^2 \Big[\lambda {\Arrowvert{\bf h}_{\mathrm{sr}}\Arrowvert^2}+(1-\lambda){\Arrowvert {\bf h}_{{\mathrm{rd}}}\Arrowvert^2}\Big]\Big),
\end{eqnarray}
where the design parameters $k_i^t$, $k_i^r>0$  for $i\in \{\mathrm{S}, \mathrm{R}, \mathrm{D}\}$ characterize the level of imperfections in the transmitter and receiver hardware, respectively. These parameters can be interpreted as the error vector magnitudes (EVMs). EVM determines  the quality of RF transceivers and is defined as the ratio of the average distortion magnitude to the average signal magnitude. Since the EVM measures the joint effect of different hardware imperfections and compensation algorithms, it can be measured directly in practice {\cite {Schenk}}. 3GPP LTE has EVM requirements in the range of $k_i^t, k_i^r \in [0.08, 0.175]$, where smaller values are needed to achieve higher spectral efficiencies {\cite {Bjor}}.

{\it Remark 1 (Co-channel Interference):} The analysis in this paper supports the scenario of large enough number of interfering signals, which is typical in wireless environments where the Gaussian assumption for the interference is valid by applying the central limit theorem {\cite {interference}}.\vspace{-4mm}

\subsection{Signal Representation}
Let us denote $x_\mathrm{S}$ and $x_\mathrm{D}$ as the unit power information signal and the jamming signal, respectively. According to the combined impact of hardware imperfections which is well-addressed by a generalized
channel model {\cite {Schenk}}, the received signal at R for DL and UL scenarios can be expressed, respectively as
\begin{eqnarray}\label{Signal_R_DL}
y_{\mathrm{R}}^\mathrm{DL}=\Big(\sqrt {\lambda P}{\bf w}_\mathrm{S}^Tx_{\mathrm{S}}+{{\boldsymbol{\eta}}_{\mathrm{S}}^{t}}^T\Big){\bf h}_\mathrm{sr}+\Big(\sqrt{(1-\lambda)P}{ x}_{\mathrm{D}}
+{{{\eta}}_{\mathrm{D}}^{t}}^T\Big){h}_\mathrm{rd}+{\eta}_{\mathrm{R}}^{r}+n_{\mathrm{R}},
\end{eqnarray}
and
\begin{eqnarray}\label{Signal_R_UL}
y_{\mathrm{R}}^\mathrm{UL}=\Big(\sqrt {\lambda P}x_\mathrm{S}+{{{\eta}}_{\mathrm{S}}^{t}}^T\Big){ h}_\mathrm{sr}+\Big(\sqrt{(1-\lambda)P}{\bf w}_\mathrm{D}^T{ x}_{\mathrm{D}}
+{{\boldsymbol {\eta}}_{\mathrm{D}}^{t}}^T\Big){\bf h}_\mathrm{rd}+{\eta}_{\mathrm{R}}^{r}+n_{\mathrm{R}},
\end{eqnarray}
where ${\bf w}_{\mathrm{S}}=\frac{{\bf h_\mathrm{sr}}^H}{{\Arrowvert {\bf h}_{{\mathrm{sr}}}\Arrowvert}}$ and ${\bf w}_{\mathrm{D}}=\frac{{\bf h_\mathrm{rd}}^H}{{\Arrowvert {\bf h}_{{\mathrm{rd}}}\Arrowvert}}$ represent the MRT transmit weight vectors at S and D, respectively. Observe from (\ref {Signal_R_DL}) and (\ref {Signal_R_UL}) that the propagated distortion noises by S and D, and the self-distortion noise at R are treated as interference at the untusted relay which is a  potential eavesdropper. As a result, the engineering insight is to beneficially forward these hardware imperfections to make the system secure instead of injecting more artificial noise by S {\cite {Khandaker1}}, {\cite {Khandaker2}}, {\cite {Lv}}, D {\cite {Mukherjee}}--{\cite {Kuhestani2}}, {\cite {error}} or a friendly jammer {\cite {Kuhestani3}}.

Then the relay amplifies its received signal in the first phase by an amplification factor of \footnote{In our analysis, we assume that the EVMs are perfectly known and will consider estimation errors in our future work.}
\begin{eqnarray}\label{gain}
G=\sqrt{\frac{P}{\mathbb{E}{|y_{\mathrm{R}}|^2}}}=\sqrt{\frac{\rho}{A_G{\lambda}+B_G}},
\end{eqnarray}
where $A_G=(\gamma_{\mathrm{sr}}-\gamma_{\mathrm{rd}})(1+{k_{{\mathrm{R}}}^{r}}^2)+{k_{{\mathrm{S}}}^{t}}^2
\gamma_{\mathrm{u}}-{k_{{\mathrm{D}}}^{t}}^2\gamma_{\mathrm{v}}$ and $B_G=\gamma_\mathrm{rd}(1+{k_{{\mathrm{R}}}^{r}}^2)+{k_{{\mathrm{D}}}^{t}}^2\gamma_{\mathrm{v}}+1$.
 Note that in the DL scenario $\gamma_{\mathrm{u}}=\rho \sum|h_{{\mathrm{sr}}_i}|^4/{\Arrowvert {\bf h}_{{\mathrm{sr}}}\Arrowvert^2}$ and $\gamma_{\mathrm{v}}=\gamma_\mathrm{rd}$, and in the UL scenario $\gamma_{\mathrm{u}}=\gamma_\mathrm{sr}$ and $\gamma_{\mathrm{v}}=\rho \sum|h_{{\mathrm{rd}}_i}|^4/{\Arrowvert {\bf h}_{{\mathrm{rd}}}\Arrowvert^2}$. Then, the received signal at D for DL and UL scenarios after self-interference (or jamming signal) cancelation are respectively, given by
\begin{align}\label{Signal_D_DL}
{\bf y}_{\mathrm{D}}^\mathrm{DL}{\hspace{-1mm}}={\hspace{-1.8mm}}\underbrace{G\sqrt {\lambda P}{\bf w}_\mathrm{S}^H{\bf h}_{{\mathrm{sr}}}{{ h}_{{\mathrm{rd}}}}{{x}_{\mathrm{S}}}}_{\mathrm{Information ~signal}}+\underbrace{G{h}_{{\mathrm{rd}}}n_{\mathrm{R}}+{n}_{\mathrm{D}}}_{\mathrm{Noise}}+\underbrace{G{{\boldsymbol{\eta}}_{\mathrm{S}}^t}^T{{\bf h}_{{\mathrm{sr}}}}{h}_{{\mathrm{rd}}}+G\eta_{\mathrm{R}}^{r}{ h}_{{\mathrm{rd}}}+G {{{\eta}}_{\mathrm{D}}^t}^T{ h}_{{\mathrm{rd}}}{{h}_{{\mathrm{dr}}}}+{{\eta}}_{\mathrm{R}}^t{ h}_{{\mathrm{rd}}}+{{\eta}}_{\mathrm{D}}^r}_{\mathrm{Distortion ~noise}},
\end{align}
and
\begin{align}\label{Signal_D_UL}
{\bf y}_{\mathrm{D}}^\mathrm{UL}=\underbrace{G\sqrt {\lambda P}{h}_{{\mathrm{sr}}}{{\bf h}_{{\mathrm{rd}}}}{{x}_{\mathrm{S}}}}_{\mathrm{Information ~signal}}+\underbrace{G{\bf h}_{{\mathrm{rd}}}n_{\mathrm{R}}+{\bf n}_{\mathrm{D}}}_{\mathrm{Noise}}+\underbrace{G{{{\eta}}_{\mathrm{S}}^t}^T{{h}_{{\mathrm{sr}}}}{\bf h}_{{\mathrm{rd}}}+G\eta_{\mathrm{R}}^{r}{\bf h}_{{\mathrm{rd}}}+G {{\boldsymbol{\eta}}_{\mathrm{D}}^t}^T{\bf h}_{{\mathrm{rd}}}{{\bf h}_{{\mathrm{dr}}}}+{{\eta}}_{\mathrm{R}}^t{\bf h}_{{\mathrm{rd}}}+{\boldsymbol{\eta}}_{\mathrm{D}}^r}_{\mathrm{Distortion ~noise}}.\nonumber\\
\end{align}
According to (\ref {Signal_R_DL}) and (\ref {Signal_R_UL}) and after some algebraic manipulations, the SNDR at R is given by
\begin{eqnarray}\label{SNR_R}
\gamma_{\mathrm{R}}=\frac{\lambda\nu}{A_\mathrm{R}{\lambda}+B_\mathrm{R}},
\end{eqnarray}
where $A_\mathrm{R}={k_{{\mathrm{R}}}^{r}}^2\nu+{k_{{\mathrm{S}}}^{t}}^2\frac{\gamma_{\mathrm{u}}}{\gamma_\mathrm{rd}}-{k_{{\mathrm{D}}}^{t}}^2
\frac{\gamma_{\mathrm{v}}}{\gamma_\mathrm{rd}}-{k_{{\mathrm{R}}}^{r}}^2-1$ and $B_\mathrm{R}=1+{k_{{\mathrm{R}}}^{r}}^2+{k_{{\mathrm{D}}}^{t}}^2
\frac{\gamma_{\mathrm{v}}}{\gamma_\mathrm{rd}}+\frac{1}{\gamma_\mathrm{rd}}$. Based on (\ref {Signal_D_DL}) and (\ref {Signal_D_UL}) and after some manipulations, the SNDR at D can be calculated as
\begin{eqnarray}\label{SNR_D}
\gamma_{\mathrm{D}}=\frac{{\lambda}\gamma_{\mathrm{sr}}}{A_\mathrm{D}{\lambda}+B_\mathrm{D}},
\end{eqnarray}
where $A_\mathrm{D}=(\gamma_{\mathrm{sr}}-{\gamma_{\mathrm{rd}}})(
{k_{{\mathrm{D}}}^{r}}^2{k_{{\mathrm{R}}}^{r}}^2+{k_{{\mathrm{R}}}^{r}}^2
{k_{{\mathrm{R}}}^{t}}^2+k_\mathrm{R}^2+{k_\mathrm{D}^r}^2)+\gamma_{\mathrm{u}}({k_{{\mathrm{D}}}^{r}}^2{k_{{\mathrm{S}}}^{t}}^2+
{k_{{\mathrm{R}}}^{t}}^2{k_{{\mathrm{S}}}^{t}}^2+{k_{{\mathrm{S}}}^{t}}^2)+\gamma_{\mathrm{v}}({k_{{\mathrm{D}}}^{r}}^2
{k_{{\mathrm{D}}}^{t}}^2-{k_{{\mathrm{R}}}^{t}}^2{k_{{\mathrm{D}}}^{t}}^2-{k_{{\mathrm{D}}}^{t}}^2)
+(\nu-1)(1+{k_{{\mathrm{R}}}^{r}}^2)
+\frac{\gamma_{\mathrm{u}}}{\gamma_\mathrm{rd}}{k_{{\mathrm{S}}}^{t}}^2-\frac{\gamma_{\mathrm{v}}}{\gamma_\mathrm{rd}}
{k_{{\mathrm{D}}}^{t}}^2$ and $B_\mathrm{D}={\gamma_{\mathrm{rd}}}({k_{{\mathrm{R}}}^{r}}^2{k_{{\mathrm{D}}}^{r}}
^2+{k_{{\mathrm{R}}}^{r}}^2{k_{{\mathrm{R}}}^{t}}^2+k_\mathrm{R}^2+{k_\mathrm{D}^r}^2)+
\gamma_{\mathrm{v}}({k_{{\mathrm{D}}}^{r}}^2{k_{{\mathrm{D}}}^{t}}^2+{k_{{\mathrm{R}}}^{t}}^2{k_{{\mathrm{D}}}^{t}}^2
+{k_{{\mathrm{D}}}^{t}}^2)+\frac{\gamma_{\mathrm{v}}}{\gamma_\mathrm{rd}}{k_{{\mathrm{D}}}^{t}}^2+\frac{1}{\gamma_\mathrm{rd}}+k_\mathrm{R}^2+{k_\mathrm{D}^r}^2+2$. We define $k_{{\mathrm{R}}}^2 {\stackrel{\tiny \Delta}{=}}{k_\mathrm{R}^{t}}^2+{k_\mathrm{R}^{r}}^2$ and $k_{{\mathrm{D}}}^2 {\stackrel{\tiny \Delta}{=}}{k_\mathrm{D}^{t}}^2+{k_\mathrm{D}^{r}}^2$ as the total imperfection level at R and D, respectively.

{\it Remark 2 (Perfect Hardware):} The received SNRs at R and D with perfect hardware were derived in {\cite {Mukherjee}}, {\cite {sun}}, {\cite {Kuhestani2}}. When setting the level of imperfections at the nodes to zero, the derived SNDRs in this section reduce to the special case as follows {\cite {sun}}
\begin{align}\label{SNR_ideal}
\gamma_\mathrm{R}^{\mathrm{perfect}}=\frac{\lambda \gamma_\mathrm{sr}}{(1-\lambda)\gamma_\mathrm{rd}+1}~~~\mathrm{and}~~~
\gamma_\mathrm{D}^{\mathrm{perfect}}=\frac{\lambda \gamma_\mathrm{sr} \gamma_\mathrm{rd}}{\lambda \gamma_\mathrm{sr} +(2-\lambda)\gamma_\mathrm{rd}+1}.
\end{align}
As can be seen, the mathematical structure of the derived SNDRs in (\ref {SNR_R}), (\ref{SNR_D}) are more complicated compared to the perfect hardware case in (\ref {SNR_ideal}), since the terms $\frac{\gamma_\mathrm{u}}{\gamma_\mathrm{rd}}$ and $\frac{\gamma_\mathrm{u}}{\gamma_\mathrm{rd}}$ manifest in the denominator. As such, it is non-trivial to propose an OPA solution for the general scenario of imperfect hardware. This generalization is done in Section III and is a main contribution of this work.

For DL scenario, (\ref{SNR_R}) and (\ref{SNR_D}) are simplified to
\begin{equation}\label{inst_sr12_L}
\gamma_\mathrm{R}=\frac{a_L \lambda}{\lambda+b_L}~~~\mathrm{and}~~~
\gamma_{\mathrm{D}}=\frac{c_L\lambda}{\lambda+d_L},
\end{equation}
where
\begin{eqnarray}\label{inst_sr12_L2}
a_L=\frac{1}{\xi_1-1},~~~b_L=\frac{\tau_1}{(\xi_1-1)\nu},~~~c_L=\frac{\gamma_\mathrm{rd}}{\tau_2 \gamma_\mathrm{rd}+\xi_1}~~~\mathrm{and}~~~d_L=\frac{\tau_3 \gamma_\mathrm{rd}+\tau_4}{\nu(\tau_2 \gamma_\mathrm{rd}+\xi_1)},
\end{eqnarray}
and, $\tau_1=1+{k_\mathrm{R}^r}^2+{k_\mathrm{D}^t}^2$, $\tau_2={k_{{\mathrm{D}}}^{r}}^2{k_{{\mathrm{R}}}^{r}}^2+
{k_{{\mathrm{R}}}^{r}}^2{k_{{\mathrm{R}}}^{t}}^2+k_\mathrm{R}^2+{k_\mathrm{D}^r}^2$,
$\tau_3=\tau_2+ {k_{{\mathrm{D}}}^{t}}^2{k_{{\mathrm{D}}}^{r}}^2+{k_{{\mathrm{R}}}^{t}}^2{k_{{\mathrm{D}}}^{t}}^2+{k_\mathrm{D}^t}^2$, $\tau_4=2+k_\mathrm{R}^2+k_\mathrm{D}^2$ and $\xi_1=1+{k_\mathrm{R}^r}^2$.
Moreover, for UL scenario, we obtain
\begin{equation}\label{inst_sr12_S}
\gamma_\mathrm{R}=\frac{a_S\lambda}{1-\lambda}~~~\mathrm{and}~~~\gamma_\mathrm{D}=\frac{b_S \lambda}{\lambda+c_S},
\end{equation}
where
\begin{eqnarray}\label{alblcl}
a_S=\frac{\nu}{\xi_1},~~~b_S=\frac{\gamma_\mathrm{sr}}{(\gamma_\mathrm{sr}-\gamma_\mathrm{rd})\tau_2+(\nu-1)\xi_1}~~~\mathrm{and}~~~
c_S=\frac{\tau_2\gamma_\mathrm{rd}+\xi_2}{(\gamma_\mathrm{sr}-\gamma_\mathrm{rd})\tau_2+(\nu-1)\xi_1},
\end{eqnarray}
and $\xi_2=2+k_\mathrm{R}^2+{k_\mathrm{D}^r}^2$. Based on (\ref {inst_sr12_L}) and (\ref {inst_sr12_S}), we can conclude that although the intercept probability is reduced by increasing the imperfection at R, the secrecy rate is also degraded. It is, therefore, of great interest to intelligently distribute the tolerable hardware imperfections across the transmission and reception radio frequency (RF) front ends of R (and other nodes) to improve the secrecy rate of the network. This hardware design approach is analyzed in Section VI and is a main contribution of this paper.

\section{Optimal Power Allocation}\label{Sec Model1}
This section proceeds to analyze the optimal power allocation problem with the aim of maximizing the instantaneous secrecy rate. Extending the results in {\cite {Kuhestani2}}, {\cite {Lv}}, {\cite {error}} where the OPA was solved for perfect hardware, we investigate the power allocation factor $\lambda$ under the presence of hardware imperfections. To do so, the instantaneous secrecy rate is evaluated by {\cite {He}}
\begin{eqnarray}\label{capacity}
R_s=\frac{1}{2 \ln 2}\Big[\ln(1+\gamma_\mathrm{D})-\ln(1+\gamma_\mathrm{R})\Big]^+.
\end{eqnarray}
By substituting $\lambda=0$ into (\ref {SNR_R}), (\ref {SNR_D}) and then (\ref {capacity}), we find $R_s=0$. Since our goal is to distribute the power optimally between S and D, a non-negative secrecy rate is achievable. As such, the instantaneous secrecy rate can be reformulated as
\begin{eqnarray}\label{inst_sr}
R_s=\frac{1}{2 \ln 2}\Big[\ln(1+\gamma_\mathrm{D})-\ln(1+\gamma_\mathrm{R})\Big].
\end{eqnarray}
Given that $\log(\cdot)$ is monotonically increasing, the maximization of $R_s$ is equivalent to the maximization of
\begin{eqnarray}\label{phi}
\phi (\lambda)~{\stackrel{\tiny \Delta}{=}}~\frac{1+\gamma_\mathrm{D}}{1+\gamma_\mathrm{R}}.
 \end{eqnarray}
Therefore, the OPA factor $\lambda^\star$ can be obtained by solving the following constrained optimization problem
\begin{eqnarray}\label{inst_sr10}
\lambda^\star={\hspace{-3mm}}&\mathrm{arg}&\mathrm{\max~}\Big\{\phi(\lambda)\Big\} \nonumber\\
 \textrm{s.t.}&&{\hspace{-5mm}}0<\lambda \leq 1
\end{eqnarray}

{\it Lemma 1}: $f(x)$ is a quasi-concave function in $\mathbb{R}$, if and only if  {\cite[Section 3.4.3] {Boyd}}
\begin{eqnarray}\label{lemma1}
\frac{\partial f(x)}{\partial x}=0 ~\Rightarrow ~\frac{\partial^2 f(x)}{\partial x^2}\leq0.
\end{eqnarray}

Based on lemma 1, we have the following corollary.\\

{\it Corollary 1}: $f(x)$ is a quasi-concave function in $x \in [x_1,x_2]$, if $\frac{\partial f(x)}{\partial x}|_{x=x_1}>0$, $\frac{\partial f(x)}{\partial x}|_{x=x_2}<0$ and there is only one maximum over  $[x_1,x_2]$ (despite constant functions).\\

{\it Proposition 1:} $\phi (\lambda)$ is a quasiconcave function of $\lambda$ in the feasible set $0< \lambda \leq 1$ and the optimal point is given by
\begin{eqnarray}\label{lambda_opt_large_nu}
\lambda^\star_E=\begin{cases}
{\frac {b_Ld_L(c_L-a_L)+ \sqrt{-a_L{b_L}c_Ld_L ({b_L}-d_L)\left( a_Ld_L-b_Lc_L-b_L +d_L\right) }}{a_Lb_L (c_L+1)-c_Ld_L(a_L+1)}}
 & ; \nu\gg 1 \\
1- \sqrt{{\frac {a_S \left( c_S+1 \right)  \left( b_S+c_S+1 \right) }{b_S c_S}}}& ; \nu\ll 1
\end{cases}
\end{eqnarray}

{\it Proof:} The first-order derivative of $\phi (\lambda)$ on $\lambda$ is given by
\begin{eqnarray}\label{derive}
\frac{\partial \phi (\lambda)}{\partial \lambda}=
\begin{cases}
{\frac { A_L {\lambda}^{2}+ B_L\lambda+C_L}{ [  \left( a_L+1 \right) \lambda+{b_L} ] ^{2}
 [ \lambda+d_L ] ^{2}}}& {\hspace{-0mm}}; \nu \gg 1 \\
{\frac { A_S{\lambda}^{2}+B_S \lambda+C_S}{ [ 1+
 \left( a_L-1 \right) \lambda ] ^{2} [ \lambda+c_L ] ^{2}
}}
& {\hspace{-0mm}}; \nu \ll 1
\end{cases},
\end{eqnarray}
where $A_L=-a_L b_L( c_L+1)+c_L d_L ( a_L+1 )$, $B_L=-2b_Ld_L ( a_L -c_L)$, $C_L= -a_L{b_L}\,{d_L}^{2}+{{b_L}}^{2}{c_L}\,d_L$, $A_S=\left( -b_L \left( a_L-1 \right) c_L-a_L \left( b_L+1 \right)  \right)$, $B_S=-2\,c_L \left( a_L+b_L \right)$ and $C_S=-a_L{c_L}^{2}+b_Lc_L$.
As can be seen from (\ref {derive}), $\frac{\partial \phi (\lambda)}{\partial \lambda}=0$ leads to two solutions on $\lambda$. It is easy to examine that the feasible solution for practical values of $k_i^t$ and $k_i^r$ {\cite {Schenk}} are derived as (\ref {lambda_opt_large_nu}). According to Corollary 1, we find that $\phi (\lambda)$ is a quasiconcave function in the feasible set.\\

To make the further analysis tractable, we provide new compact expressions for the OPA in the high SNR regime. In the case of DL, the expressions in (\ref {inst_sr12_L2}) are simplified to
\begin{eqnarray}\label{as_abcd}
a_L=\frac{1}{\xi_1-1},~~b_L=\frac{\tau_1}{(\xi_1-1)\nu},~~c_L=\frac{1}{\tau_2 },~~d_L=\frac{\tau_3 }{\tau_2 \nu},
\end{eqnarray}
and in the case of UL scenario, the expressions in (\ref {alblcl}) are changed to
\begin{eqnarray}\label{as_abc}
a_S=\frac{\nu}{\xi_1},~~b_S=\frac{\nu}{(\nu-1)\tau_2},~~
c_S=\frac{1}{\nu-1}.
\end{eqnarray}
By substituting (\ref {as_abcd}) and (\ref {as_abc}) into (\ref {lambda_opt_large_nu}), the OPA solutions in the high SNR regime can be expressed in the following tractable forms
\begin{eqnarray}\label{lambda_sol_H}
\lambda^\star_{{High}}{\hspace{-1mm}}= {\hspace{-1mm}}\begin{cases}
\frac{\theta_L}{\nu} & ; \mathrm{DL} \\
1-{\theta_S}\nu& ; \mathrm{UL}
\end{cases}
\end{eqnarray}
where $\theta_L=\sqrt{\frac{\tau_3}{\tau_2}(\tau_1-\tau_3)}+\frac{\tau_3}{\tau_2}(\xi_1-1)-\tau_3$ and $\theta_S=\sqrt{\frac{1+\tau_2}{\xi_1}}$.

\section{Ergodic Secrecy Rate}
In this section, we derive the ESR  of the proposed secure transmission scheme in each case of DL and UL scenarios. Since it is not straightforward to obtain a closed-form expression for the exact ESR of DL and UL scenarios (the exact ESR includes double integral expressions due to the complicated structures of (\ref {lambda_opt_large_nu})), we therefore proceed by first deriving new analytical expressions for the ESR in the high SNR regime that can be applied to arbitrary channel fading distributions. Based on these, new closed-form expressions are derived for the ESR in Rayleigh fading channels. Despite prior works in the literature {\cite {sun}}, {\cite {Asma}}--{\cite {sec-aware}} that investigated the ESR based on perfect hardware assumption in various untrusted relaying networks, we take into account hardware imperfections. The new results in this section generalize the recent results in {\cite {sun}}, {\cite {Kuhestani2}}.

The ESR as a useful secrecy metric representing the rate below which any average secure transmission rate is achievable {\cite {review3}}. As such, the ESR is given by
\begin{eqnarray}\label{Total}
\overline{R_s}=\mathbb{E}\Big\{R_s\Big\}=\frac{1}{2 \ln 2} \Big[\underbrace{\mathbb{E}\Big\{\ln(1+\gamma_\mathrm{D})\Big\}}_{T_1}-\underbrace{\mathbb{E}\Big\{\ln(1+\gamma_\mathrm{R})\Big\}}_{T_2}\Big].
\end{eqnarray}
In the following, we proceed to evaluate the parts $T_1$ and $T_2$ and then $\overline{R_s}$ for each case of DL and UL scenarios. Towards this goal, we present the following useful lemma.\\

{\it Lemma 2:} For positive constants $\alpha_1$,  $\alpha_2$ and  $\alpha_3$, and non-negative r.v. $\Gamma$, the cdf of the new r.v.  $\widehat{\Gamma}=\frac{\alpha_1 \Gamma}{\alpha_2 \Gamma + \alpha_3}$ is derived as
\begin{eqnarray}
F_{\widehat{\Gamma}}(x)=\begin{cases}F_{\Gamma}\Big(\frac{\alpha_3 x}{\alpha_1-\alpha_2 x}\Big)&; 0 \leq x < \frac{\alpha_1}{\alpha_2} \\
1 &; x \geq \frac{\alpha_1}{\alpha_2}
\end{cases}
\end{eqnarray}

{\it Proof:} We start from the definition of the cdf as follows
\begin{eqnarray}
F_{\widehat{\Gamma}}(x)=\Pr \Big\{\frac{\alpha_1 \Gamma}{\alpha_2 \Gamma + \alpha_3}\leq x\Big\}=\Pr \Big\{\Gamma (\alpha_1-\alpha_2 x)\leq \alpha_3 x\Big\},
\end{eqnarray}
where the last probability equals to one for $\alpha_1-\alpha_2 x < 0$. Otherwise, it equals to $F_{\Gamma}\Big(\frac{\alpha_3 x}{\alpha_1-\alpha_2 x}\Big)$. \\

{\it 1) Downlink Scenario:} By plugging (\ref {lambda_sol_H}) into (\ref {inst_sr12_L}), we obtain
\begin{eqnarray}\label{grd_H}
\gamma_\mathrm{R}{\hspace{-2mm}}&=&{\hspace{-2mm}}\frac{\theta_L}{\theta_L (\xi_1-1)+\tau_1},\\
\gamma_\mathrm{D}{\hspace{-2mm}}&=&{\hspace{-2mm}}\frac{\theta_L \gamma_\mathrm{rd}}{ (\tau_2 \theta_L+\tau_3) \gamma_\mathrm{rd}+\xi_1 \theta_L +\tau_4}.\label{grd_H2}\end{eqnarray}
We find that all the terms in (\ref {grd_H}) and (\ref {grd_H2}) are deterministic constants which leads to the secrecy rate ceiling in the high SNR regime.

Based on lemma 2 and (\ref {grd_H2}),  the part $T_1$ in (\ref {Total}) is given by
\begin{eqnarray}\label{T_1_DL}
T_1=\mathbb{E}\Big\{\ln\Big(1+\frac{\theta_L \gamma_\mathrm{rd}}{ (\tau_2 \theta_L+\tau_3) \gamma_\mathrm{rd}+\xi_1 \theta_L +\tau_4}\Big)\Big\}=\int_0^{\frac{\theta_L}{\tau_2 \theta_L+\tau_3}}\frac{1-F_{\gamma_\mathrm{rd}}\Big(\frac{(\xi_1 \theta_L+\tau_4)x}{\theta_L-(\tau_2\theta_L+\tau_3)x}\Big)}{1+x}\mathrm{d}x,
\end{eqnarray}
where the last equation follows from the integration by parts. The  expression in (\ref {T_1_DL}) is straightforwardly evaluated for any channel fading distribution, either directly or by a simple numerical integration. \\Furthermore, based on (\ref {grd_H2}) the part $T_2$ is a constant value as
\begin{eqnarray}\label{T_2_DL}
T_2=\ln \Big(1+\frac{\theta_L}{\theta_L (\xi_1-1)+\tau_1}\Big).
\end{eqnarray}
We conclude from (\ref {T_2_DL}) that the amount of information leakage is independent of the transmit SNR and the position of the relay, and only depends on the EVMs at network nodes. By replacing (\ref {T_1_DL}) and (\ref {T_2_DL}) into (\ref {Total}), the compact ESR expression is achieved for any channel distribution.

For the case of Rayleigh fading, due to the fact that $\gamma_\mathrm{rd}$ is an exponential r.v. and applying {\cite[Eq. (4.337.2)]{table}}, the part $T_1$ can be expressed in a closed-form solution. By substituting this and (\ref {T_2_DL}) into (\ref {Total}), the closed-form ESR expression becomes
\begin{align}\label{Total_Rate_DL2}
\overline{R_s}^{\mathrm{DL}}=\frac{1}{2 \ln 2} \Big[e^{\frac{1}{r_2 \overline{\gamma}_\mathrm{rd}}}\mathrm{Ei}(-\frac{1}{r_2 \overline{\gamma}_\mathrm{rd}})-e^{\frac{1}{r_1 \overline{\gamma}_\mathrm{rd}}}\mathrm{Ei}(-\frac{1}{r_1 \overline{\gamma}_\mathrm{rd}})-\ln \Big(1+\frac{\theta_L}{\theta_L (\xi_1-1)+\tau_1}\Big)\Big],
\end{align}
where $r_1=\frac{(1+\tau_2)\theta_L+\tau_3}{\xi_1 \theta_L +\tau_4}$ and $r_2=\frac{\tau_2\theta_L+\tau_3}{\xi_1 \theta_L +\tau_4}$.{\hspace{-.5mm}} We conclude from \eqref{Total_Rate_DL2} that the ESR is exclusively characterized by the level of imperfections over nodes and $\overline{\gamma}_{\mathrm{rd}}$ which is a function of the transmit SNR and the distance-dependent channel gain $\mu_{\mathrm{rd}}$.\\

{\it 2) Uplink Scenario:} Substituting (\ref {lambda_sol_H}) into (\ref {inst_sr12_S}) yields
\begin{eqnarray}\label{grd_L}
\gamma_\mathrm{R}{\hspace{-3mm}}&=&{\hspace{-3mm}}\frac{\lambda^\star_{\mathrm{H}}\nu}{(1-\lambda^\star_{\mathrm{H}})\xi_1}\approx \frac{1}{\sqrt{\xi_1 (1+\tau_2)}},\\
\gamma_\mathrm{D}{\hspace{-3mm}}&\approx&{\hspace{-3mm}}\frac{\gamma_\mathrm{sr}}{\tau_2(1+\sqrt{\frac{1+\tau_2}{\xi_1}})\gamma_\mathrm{sr}+\xi_2-\xi_1}.
\label{grd_L2}\end{eqnarray}
Observe from (\ref {grd_L}) and (\ref {grd_L2}) that only the first hop SNR, $\gamma_\mathrm{sr}$ contributes to the secrecy rate performance. Following the similar procedure as the DL scenario, the ESR performance of the UL case for arbitrary fading distribution can be expressed as
\begin{align}\label{rate_S1}
\overline{R_s}^{\mathrm{UL}}=\frac{1}{2 \ln 2}\Big(\int_0^{\frac{1}{\tau_2(1+\sqrt{\frac{1+\tau_2}{\xi_1}})}}\frac{1-F_{\gamma_\mathrm{sr}}(\frac{(\xi_2-\xi_1)x}{1-\tau_2
(1+\sqrt{\frac{1+\tau_2}{\xi_1}})x})}{1+x}~\mathrm{d}x-\ln (1+\frac{1}{\sqrt{\xi_1 (1+\tau_2)}})\Big).
\end{align}

For the case of Rayleigh fading, the closed-form ESR expression is given by
\begin{align}\label{Total_Rate_UL}
\overline{R_s}^{\mathrm{UL}}=\frac{1}{2 \ln 2} \Big[e^{\frac{1}{t_2 \overline{\gamma}_\mathrm{sr}}}\mathrm{Ei}(-\frac{1}{t_2 \overline{\gamma}_\mathrm{sr}})-e^{\frac{1}{t_1 \overline{\gamma}_\mathrm{sr}}}\mathrm{Ei}(-\frac{1}{t_1 \overline{\gamma}_\mathrm{sr}})-\ln \Big(1+\frac{1}{\sqrt{\xi_1 (1+\tau_2)}}\Big)\Big],
\end{align}
where $t_1=\frac{1+\tau_2(1+\sqrt{\frac{1+\tau_2}{\xi_1}})}{\xi_2-\xi_1}$ and $t_2=\frac{\tau_2(1+\sqrt{\frac{1+\tau_2}{\xi_1}})}{\xi_2-\xi_1}$.
It is observed from \eqref{Total_Rate_UL} that the ESR  is entirely determined by the average channel gain of the first
hop, the transmit SNR and the level of imperfections of all the network nodes. We also find that increasing the number of antennas at D has no impact on the ESR when $N_\mathrm{d}$ is large.
{\vspace{-.4mm}}

\section{Secrecy Outage Probability}
In this section, similar to our ESR results, general expressions are first presented for the SOP that can be applied to any channel distribution, under the presence of transceiver imperfections and in the high SNR regime. Based on these, we derive novel closed-form expressions for the SOP in Rayleigh fading channels.

The SOP denoted by $P_\mathrm{so}$ is a criterion that determines the fraction of fading realizations where a secrecy rate $R_t$ cannot be supported {\cite {Mukherjee}}. Accordingly, the overall SOP is defined as the probability that a system with the instantaneous secrecy rate $R_s$ is not able to support the target transmission rate $R_t$; $P_{\mathrm{so}}=\Pr\Big\{R_s<R_t\Big\}.$

In the following, we focus on each case of DL and UL, respectively.\\

{\it 1) Downlink Scenario:} Substituting (\ref {grd_H2}) into (\ref {inst_sr}) and then based on the SOP definition we obtain
\begin{align}\label{SOP_DL_general}
P_\mathrm{so}^\mathrm{DL}&=\Pr \Big(\frac{\theta_L \gamma_\mathrm{rd}}{ (\tau_2 \theta_L+\tau_3) \gamma_\mathrm{rd}+\xi_1 \theta_L +\tau_4}\leq \widetilde{R_t}\Big)\nonumber\\
&=\begin{cases}
F_{\gamma_\mathrm{rd}}\Big(\frac{(\xi_1 \theta_L+\tau_4)\widetilde{R_t}}{\theta_L-(\tau_2 \theta_L+\tau_3)\widetilde{R_t}}\Big) & ; R_t < \frac{1}{2} \log_2 \Big( \frac{1+\frac{\theta_L}{\tau_2 \theta_L+\tau_3}}{1+\gamma_\mathrm{R}}\Big) \\
1& ; R_t \geq \frac{1}{2} \log_2 \Big( \frac{1+\frac{\theta_L}{\tau_2 \theta_L+\tau_3}}{1+\gamma_\mathrm{R}}\Big)
\end{cases}
\end{align}
where $\widetilde{R_t}=2^{2R_t}(1+\gamma_\mathrm{R})-1$ and  $\gamma_\mathrm{R}$ is in (\ref {grd_H}), and the last equation follows from using lemma 2. It is worth pointing out that the SOP expressions in (\ref {SOP_DL_general}) allows the straightforward evaluation of the SOP for any channel fading distribution by a simple numerical integration. We can conclude from (\ref {SOP_DL_general}) that the SOP is always 1 for target transmission rates more than a threshold (which only depends on the EVMs of the nodes). Interestingly, this event holds for any channel fading distribution, any network topology and any transmit SNR. Therefore as explained in Section IV, some secrecy rates can never be achieved due to secrecy rate ceiling. Furthermore, we conclude that for target transmission rates smaller than the threshold, $P_\mathrm{so}$ approaches zero with increasing SNR (similar to perfect hardware) whereas the SOP always equals one for target transmission rates larger than the threshold. This result is fundamentally different to the perfect hardware case where the SOP goes to zero with increasing SNR and for any target transmission rate {\cite {Mukherjee}}, {\cite {Teh1}}, {\cite {Kuhestani2}}.

For Rayleigh fading channels, $\gamma_\mathrm{rd}$ is an exponential r.v. and therefore, our new and simple
closed-form SOP expression in the presence of transceiver hardware imperfection is given by

\begin{align}\label{SOP_DL_rayleigh}
P_\mathrm{so}^\mathrm{DL}=
\begin{cases}
1-\exp\Big(-\frac{(\xi_1 \theta_L+\tau_4)\widetilde{R_t}}{(\theta_L-(\tau_2 \theta_L+\tau_3)\widetilde{R_t})\overline{\gamma}_\mathrm{rd}}\Big) & ;  R_t < \frac{1}{2} \log_2 \Big( \frac{1+\frac{\theta_L}{\tau_2 \theta_L+\tau_3}}{1+\gamma_\mathrm{R}}\Big) \\
1& ; R_t \geq  \frac{1}{2} \log_2 \Big( \frac{1+\frac{\theta_L}{\tau_2 \theta_L+\tau_3}}{1+\gamma_\mathrm{R}}\Big)
\end{cases}
\end{align}
We note that the results of this section generalize the results of {\cite {Kuhestani2}} which were derived for the case of untrusted relaying with perfect hardware.\\

{\it 2) Uplink Scenario:} Similar to the DL case, the SOP can be obtained by substituting $\gamma_\mathrm{D}$ in (\ref {grd_L2}) into (\ref {inst_sr}) and using the SOP definition. This yields
\begin{align}\label{SOP_UL_general}
P_\mathrm{so}^\mathrm{UL}=\begin{cases}
F_{\gamma_\mathrm{sr}}\Big(\frac{(\xi_2-\xi_1)\widetilde{R_t}}
{1-\tau_2(1+\sqrt{\frac{1+\tau_2}{\xi_1}})\widetilde{R_t}}\Big) & ; R_t < \frac{1}{2} \log_2 \Big( \frac{1+\frac{1}{\tau_2(1+\sqrt{\frac{1+\tau_2}{\xi_1}})}}{1+\gamma_\mathrm{R}}\Big) \\
1& ; R_t \geq {\hspace{-.7mm}} \frac{1}{2} \log_2 \Big( \frac{1+\frac{1}{\tau_2(1+\sqrt{\frac{1+\tau_2}{\xi_1}})}}{1+\gamma_\mathrm{R}}\Big)
\end{cases}
\end{align}
where $\widetilde{R_t}=2^{2R_t}(1+\gamma_\mathrm{R})-1$ and  $\gamma_\mathrm{R}$ is in (\ref {grd_L}). For the special case of Rayleigh fading channels, the closed-form SOP is derived as
\begin{align}\label{SOP_UL_rayleigh}
P_\mathrm{so}^\mathrm{UL}=\begin{cases}
1-\exp{\hspace{-1mm}}\Big(\frac{-(\xi_2-\xi_1)\widetilde{R_t}}
{(1-\tau_2(1+\sqrt{\frac{1+\tau_2}{\xi_1}})\widetilde{R_t})\overline{\gamma}_\mathrm{sr}}\Big) & ; R_t  < \frac{1}{2} \log_2 \Big( \frac{1+\frac{1}{\tau_2(1+\sqrt{\frac{1+\tau_2}{\xi_1}})}}{1+\gamma_\mathrm{R}}\Big) \\
1& ; R_t \geq \frac{1}{2} \log_2 \Big( \frac{1+\frac{1}{\tau_2(1+\sqrt{\frac{1+\tau_2}{\xi_1}})}}{1+\gamma_\mathrm{R}}\Big)
\end{cases}
\end{align}
As observed in the numerical results, the closed-form expressions (\ref {SOP_DL_rayleigh}) and (\ref {SOP_UL_rayleigh}) are sufficiently tight at medium and high transmit SNRs.

\section{Hardware Design}

In this section, we aim to gain some insights into the design of RF hardware to maximize the secrecy rate in untrusted relaying networks. Depending on the fixed cost of each network node, we show how the RF segments at the transmission and reception front ends can be designed. Specially, given the maximum tolerable hardware imperfection of each node, we derive new analytical results characterizing how the hardware imperfections should be distributed between the transmission RF segment and the reception RF segment of each node to maximize the secrecy rate. Therefore, we should find $k_i^t$ and $k_i^r$, $i \in \{\mathrm{S}, \mathrm{R}, \mathrm{D}\}$ to maximize the secrecy rate such that $k_i^t+k_i^r=k_i^\mathrm{tot}$. Mathematically speaking, our goal is to solve the following optimization problem
\begin{align} \label{optimization_ptoblem}
(k_\mathrm{R}^t, k_\mathrm{R}^r, k_\mathrm{D}^t, k_\mathrm{D}^r)&= \mathrm{arg} ~~\max \phi (\lambda^{\star})\\
\mathrm{s.t.}& {\hspace{2mm}}k_\mathrm{R}^t+k_\mathrm{R}^r = k_\mathrm{R}^\mathrm{tot}\nonumber\\
& {\hspace{2mm}} k_\mathrm{D}^t+k_\mathrm{D}^r = k_\mathrm{D}^\mathrm{tot}\nonumber
\end{align}
Based on (\ref {grd_H2}), (\ref {grd_L2}) and (\ref {inst_sr}), the instantaneous secrecy rate is an increasing function of the transmit SNR. Since it is our aim to achieve high transmission rates, we consider the asymptotic SNR regime $\rho \rightarrow \infty$ {\cite {Schenk}} to solve the hardware design problem (\ref {optimization_ptoblem}). As observed in numerical studies, the results of the high SNR analysis can be applied successfully at finite SNRs.

In the asymptotic SNR regime and for any random distributions on $\gamma_\mathrm{sr}$ and $\gamma_\mathrm{rd}$, the asymptotic received SNDRs at R and D are respectively, given by
\begin{eqnarray}\label{asymp_R}
\gamma_\mathrm{R}^{\infty}=\begin{cases}
\frac{\theta_L}{\theta_L (\xi_1-1)+\tau_1}& ; \mathrm{DL} \\
\frac{1}{\sqrt{\xi_1(1+\tau_2)}}& ; \mathrm{UL}
\end{cases},
\end{eqnarray}
and
\begin{eqnarray}\label{asymp_D}
\gamma_\mathrm{D}^{\infty}=\begin{cases}
\frac{\theta_L}{\tau_2 \theta_L+\tau_3}& ; \mathrm{DL} \\
\frac{1}{\tau_2 (1+\sqrt{\frac{1+\tau_2}{\xi_1}})}& ; \mathrm{UL}
\end{cases}.
\end{eqnarray}
By substituting (\ref {asymp_R}) and (\ref {asymp_D}) into (\ref {phi}), the secrecy rate ceiling is given by
\begin{eqnarray}\label{phi_infty}
\phi^{\infty}=\begin{cases}
\frac{((1+\tau_2)\theta_L+\tau_3)((\xi_1-1)\theta_L+\tau_1)}{(\xi_1 \theta_L+\tau_1) (\tau_2 \theta_L+\tau_3)}& ; \mathrm{DL} \\
{\frac {\sqrt {\xi_1} (\tau_2+1)\left( \tau_2+\sqrt{\xi_1}\sqrt {\tau_2+1} \right)
}{\tau_2 \left(\sqrt{ \xi_1}+ \sqrt {\tau_2+1}
 \right)  \left( \sqrt {\xi_1}\sqrt {\tau_2+1}+1 \right) }}
& ; \mathrm{UL}
\end{cases}
 \end{eqnarray}

Some conclusions and insights can be concluded from (\ref {phi_infty}). First, the secrecy rate ceiling event appears in the asymptotic SNR regime, which significantly limits the performance of the system. This event is different from the perfect hardware case, in which the ESR increases with increasing SNR. Note that this ceiling effect is independent of the fading distribution.

In the following, we focus on the of DL and UL scenarios separately and then conclude about the hardware design of the overall network.
{\vspace{-1mm}}

\subsection{Downlink Scenario:}
{\vspace{-1mm}}
In the following, we proceed to solve the optimization problem (\ref {optimization_ptoblem}) by independently discussing on the hardware design at R and D as follows.\\

{\it Proposition 2:} Suppose $k_\mathrm{R}^t+k_\mathrm{R}^r=k_\mathrm{R}^{\mathrm{tot}}$, hence the secrecy rate ceiling is maximized if $k_\mathrm{R}^t=k_\mathrm{R}^r=\frac{k_\mathrm{R}^{\mathrm{tot}}}{2}$.

{\it Proof:}  Please see Appendix \ref{appA}.\\

{\it Proposition 3:} Suppose $k_\mathrm{D}^t+k_\mathrm{D}^r=k_\mathrm{D}^{\mathrm{tot}}$, thus the secrecy rate ceiling is maximized if
\begin{align}\label{Prop3}
k_\mathrm{D}^t{\hspace{-1mm}}={\hspace{-1mm}}\frac{2k_\mathrm{R}^2{\hspace{-1mm}}+{\hspace{-1mm}}2{k_\mathrm{D}^\mathrm{tot}}^2
{\hspace{-1.5mm}}+{\hspace{-1mm}}3
{\hspace{-1mm}}-{\hspace{-1mm}}\sqrt{4k_\mathrm{R}^4{\hspace{-1.2mm}}+{\hspace{-.8mm}}8k_\mathrm{R}^2{k_\mathrm{D}^\mathrm{tot}}^2
{\hspace{-1.2mm}}+{\hspace{-.8mm}}4{k_\mathrm{D}^\mathrm{tot}}^4{\hspace{-1.4mm}}+{\hspace{-.8mm}}12k_\mathrm{R}^2{\hspace{-1mm}}-{\hspace{-1mm}}
4{k_\mathrm{D}^\mathrm{tot}}^2{\hspace{-1.4mm}}+{\hspace{-.8mm}}9}}{4k_\mathrm{D}^\mathrm{tot}}.
\end{align}
{\it Proof:} Please see Appendix \ref{appB}.\\

\subsection{Uplink Scenario:}
Similar to DL scenario, two propositions are provided as follows.\\

{\it Proposition 4:} Suppose $k_\mathrm{R}^t+k_\mathrm{R}^r=k_\mathrm{R}^{\mathrm{tot}}$, thus the secrecy rate ceiling is maximized if $k_\mathrm{R}^t=k_\mathrm{R}^r=\frac{k_\mathrm{R}^{\mathrm{tot}}}{2}$.

{\it Proof:}  Please see Appendix \ref{appC}.\\

{\it Proposition 5:} Suppose $k_\mathrm{D}^t+k_\mathrm{D}^r=k_\mathrm{D}^{\mathrm{tot}}$, hence the secrecy rate ceiling is a monotonically decreasing function of $k_\mathrm{D}^r$.\\

{\it Proof:}  In this case, we have
\begin{eqnarray} \label{small_der_phi_tot}
\frac{\partial \phi^{\infty} }{\partial k_\mathrm{D}^r}=\frac{\partial \phi^{\infty}}{\partial \tau_2 } \frac{\partial \tau_2}{\partial k_\mathrm{D}^r},
\end{eqnarray}
where $\frac{\partial \tau_2}{\partial k_\mathrm{D}^r}=2{k_\mathrm{R}^r}^2k_\mathrm{D}^r +2 k_\mathrm{D}^r>0$ and $\frac{\partial \phi^{\infty}}{\partial \tau_2 }$ in (\ref {small_der_phi_1}) is negative in the feasible set. As such, $\frac{\partial \phi^{\infty} }{\partial k_\mathrm{D}^r} <0$.\\

Based on Propositions 2--5, we provide the following corollary as a conclusion of the analysis which provides new insights into the system design.\\

{\it Corollary 2:}  Consider a cooperative network in which one multiple antennas node communicates with a single antenna node via a single antenna untrusted relay. Let us assume a predefined cost can be assigned to each node. To maximize the secrecy rate of this network the following considerations should be taken into account:
\begin{itemize}
  \item According to Propositions 2 and 4, the total cost for the relay node should be divided by half between the transmission and reception RF front ends, i,e., it is better to apply the same level of imperfections at every transceiver chain, instead of utilizing a mix of high-quality and low-quality transceiver chains.
  \item According to Proposition 5, to design the multiple antennas node, the designers are persuaded to use higher-quality hardware in reception RF front end and lower-quality hardware in the transmission RF front end, i.e, the hardware imperfections at the reception end of the multiple antennas node should be close to zero.
        \item According to Proposition 3, to design the single antenna node, the quality of RF requirements at the transmission end should obeys from (\ref {Prop3}). As observed in the numerical examples, we obtain $k_\mathrm{D}^t>\frac{k_\mathrm{D}^\mathrm{tot}}{2}$ for typical values of EVMs.
\end{itemize}

\begin{figure}[t]
  \begin{center}{\hspace{-1mm}}
    \includegraphics[width=3.2in,height=3.4in]{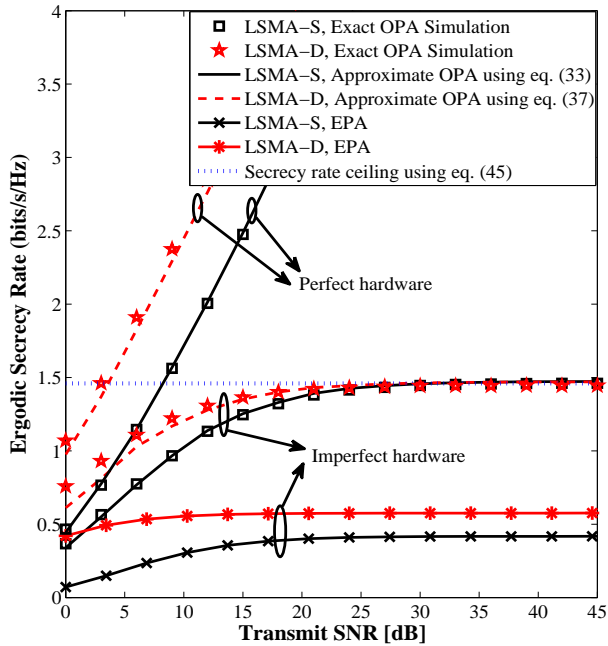} \vspace{3mm}
    \caption{Ergodic secrecy rate versus transmit SNR for exact and the derived closed-form expressions under perfect and imperfect transceiver hardwares. Number of antennas at source or destination is set to 16. For imperfect case with $k=0.1$, the secrecy rate ceiling is observed.}
  \label{esr_16}\end{center}
\end{figure}

\begin{figure}[t]
  \begin{center}
    \includegraphics[width=3.2in,height=3.4in]{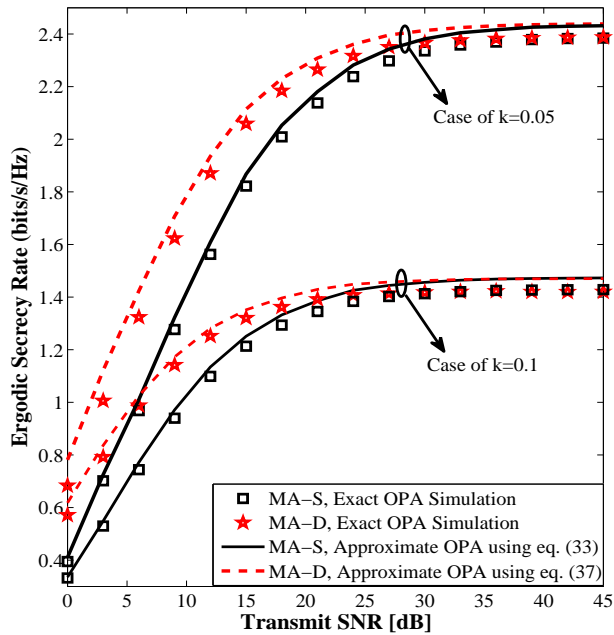} 
    \caption{Ergodic secrecy rate versus transmit SNR for exact and the derived closed-form expressions under different levels of hardware imperfections; $k \in \{0.05, 0.1\}$. Number of antennas at source or destination is set to 4.}
  \label{esr_4}\end{center}
\end{figure}
\begin{figure}[t]
\begin{center}
    \includegraphics[width=3.4in,height=3.4in]{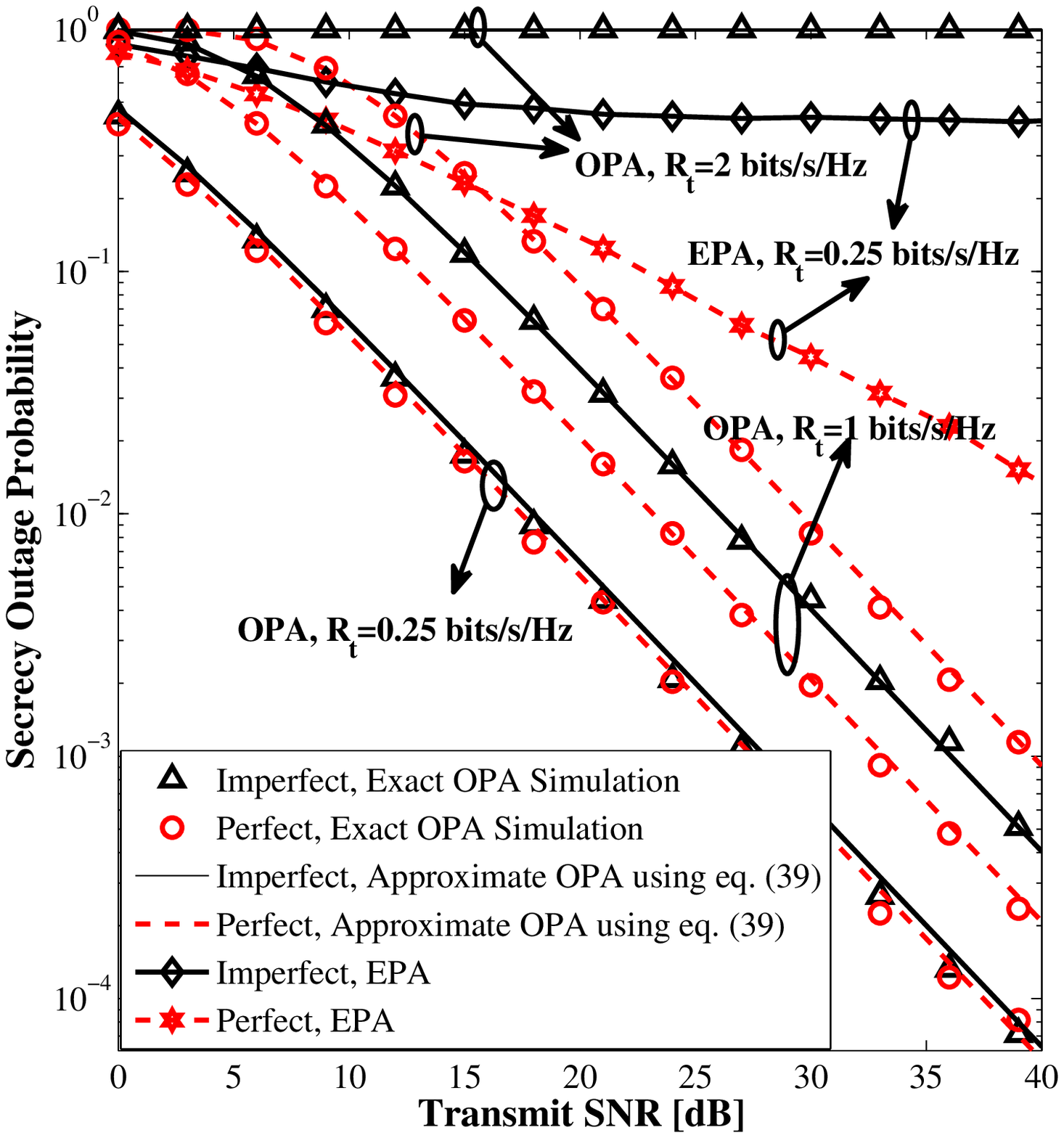} 
    \caption{Secrecy outage probability versus transmit SNR for DL transmission, with different target transmission rates and under perfect $(k=0)$ and imperfect hardware $(k=0.1)$.}
  \label{sop_dl}\end{center}
\end{figure}

\begin{figure}[t]
  \begin{center}{\hspace{-1mm}}
    \includegraphics[width=3.2in,height=3.4in]{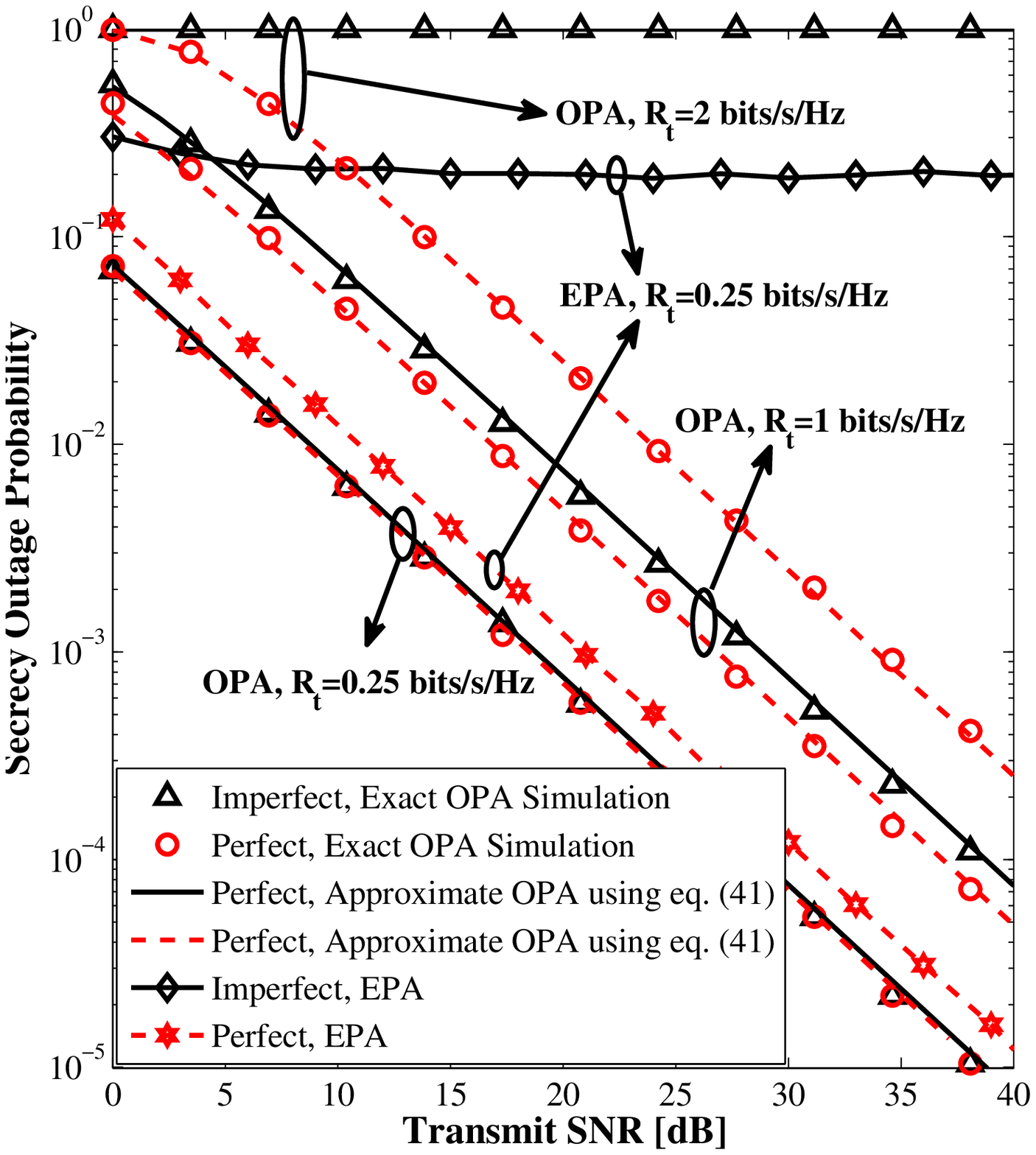} 
    \caption{Secrecy outage probability versus transmit SNR for UL transmission, with different target transmission rates and under perfect $(k=0)$ and imperfect hardware $(k=0.1)$.}
  \label{sop_ul}\end{center}
\end{figure}

\begin{figure}[t] \label{1}
  \begin{center}{\hspace{-1mm}}
    \includegraphics[width=3.2in,height=3.4in]{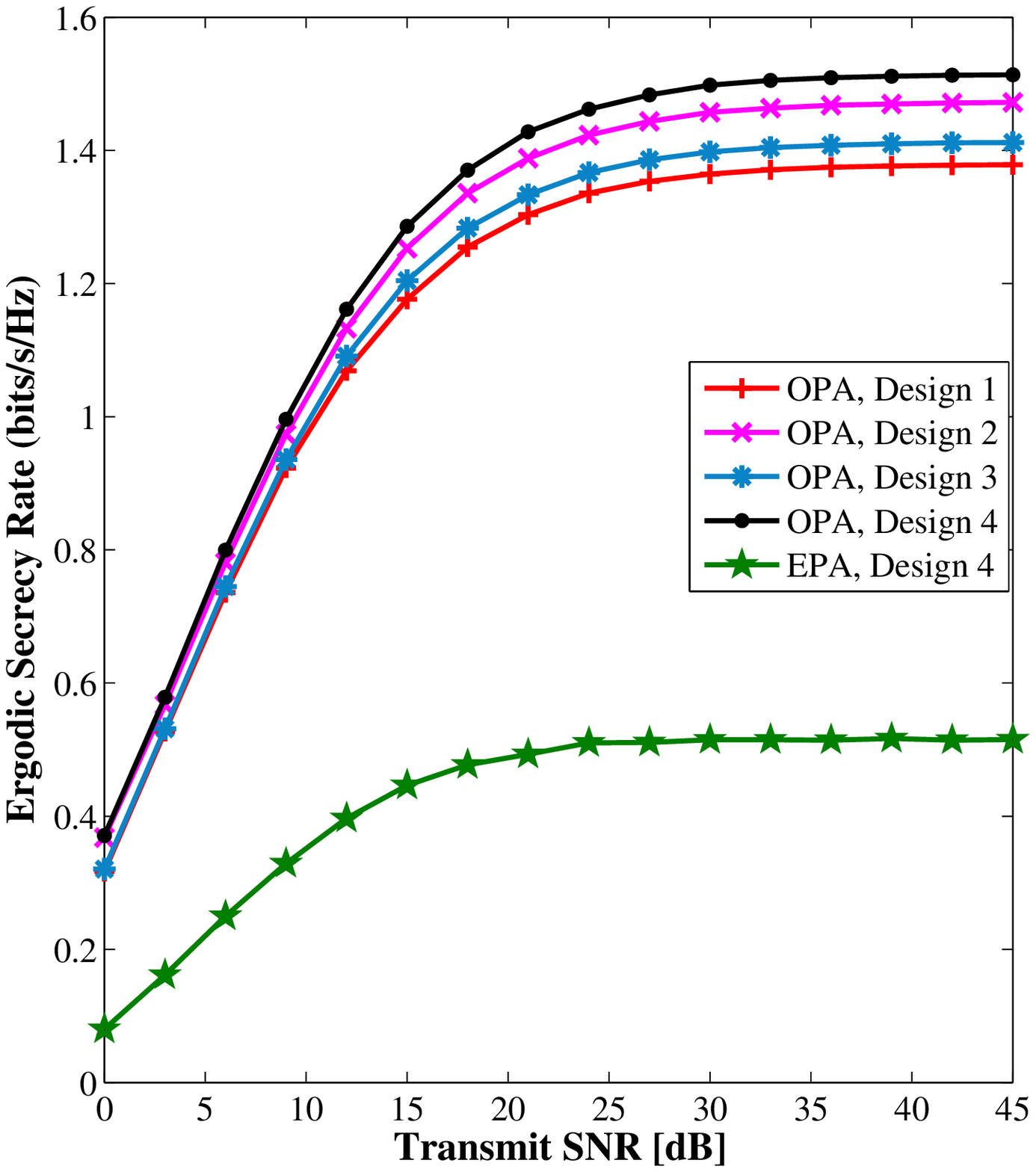} 
    \caption{Ergodic secrecy rate versus transmit SNR for LSMA at S. Various imperfection distributions over RF transmission and reception ends are considered  for $k_\mathrm{R}^\mathrm{tot}=k_\mathrm{D}^\mathrm{tot}=0.2$.}
  \label{design_dl}\end{center}
\end{figure}

\begin{figure}[t] \label{1}
  \begin{center}{\hspace{-1mm}}
    \includegraphics[width=3.2in,height=3.4in]{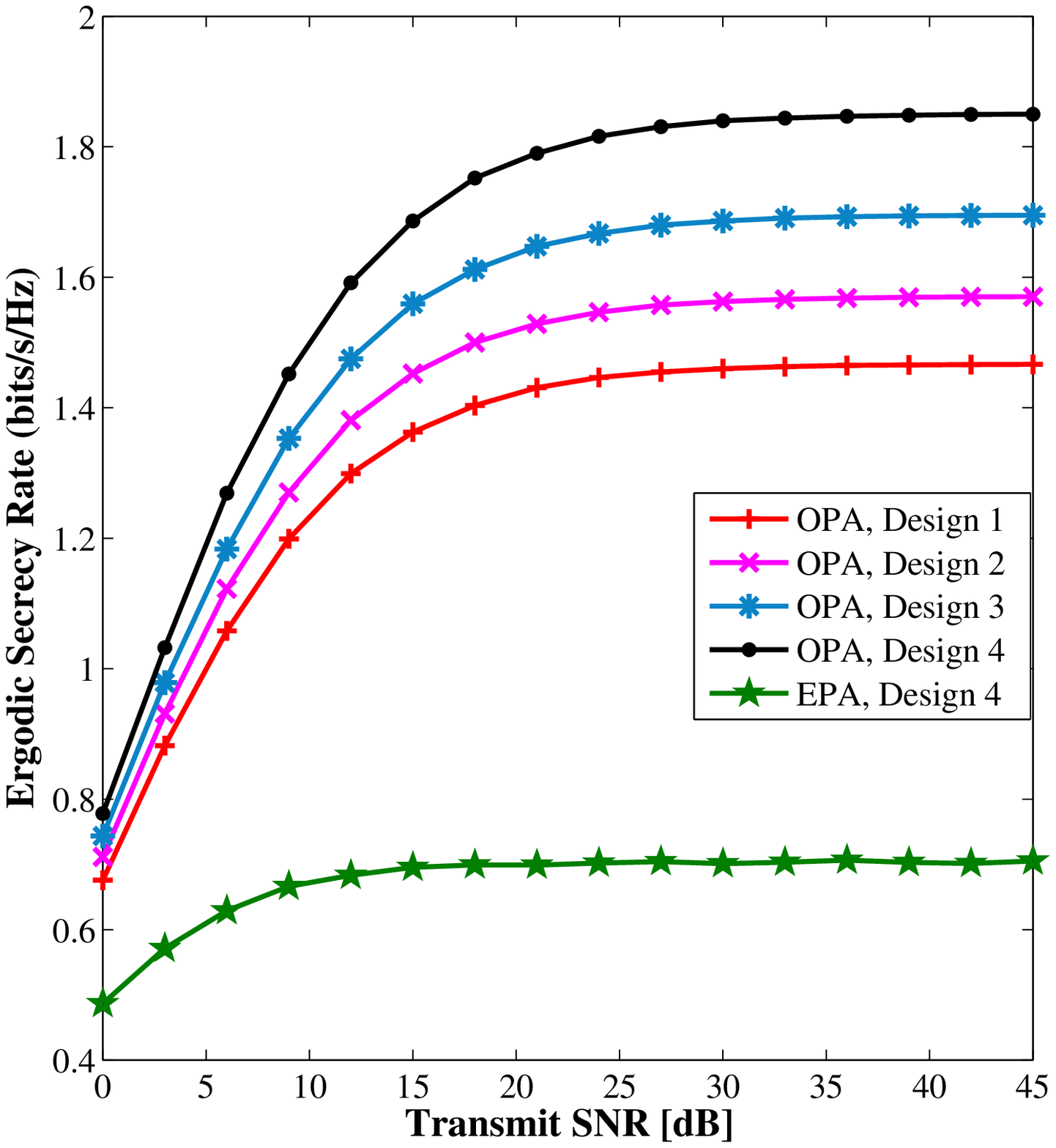} 
    \caption{Ergodic secrecy rate versus transmit SNR for LSMA at D. Various imperfection distributions over RF transmission and reception ends are considered for $k_\mathrm{R}^\mathrm{tot}=k_\mathrm{D}^\mathrm{tot}=0.2$.}
  \label{design_ul}\end{center}
\end{figure}

\section{Numerical Results and Discussions}
In this section, numerical results are provided to verify the accuracy of the derived closed-form expressions in Section IV and V for LSMA at S (LSMA-S) and LSMA at D (LSMA-D), respectively, and also the cases of multiple antennas at S (MA-S) and multiple antennas at D (MA-D). We compare our LSMA-based ESR performance with the exact ESR with Monte-Carlo simulations where the OPA is numerically evaluated for finite numbers of antennas using the bisection method. In addition, the equal power allocation (EPA) between S and D (i.e., $\lambda=0.5$) is plotted as a benchmark. Furthermore, the concepts of secrecy rate ceiling and the practical hardware insights from Section VI are numerically presented. In our numerical evaluations, the transmission links between nodes are modeled by the Rayleigh fading channel and
the average channel gains are specified as $\mu_\mathrm{sr}=\mu_\mathrm{rd}=10$. Moreover, for LSMA the number of antennas is set to 16, and for MA the number of antennas is set to 4.

Fig. \ref {esr_16} depicts the ESR versus transmit SNR $\rho$ in dB for both cases of DL and UL and for perfect ($k=k_\mathrm{R}^t=k_\mathrm{R}^r=k_\mathrm{S}^t=k_\mathrm{D}^t=k_\mathrm{D}^r=0$) and imperfect ($k=0.1$) cases. The number of antennas at S and D are set to $N_\mathrm{s}=N_\mathrm{d}=16$. It is observed from the figure that the Monte-Carlo simulation of the exact OPA evaluated using the bisection method is in good agreement with the derived high SNR closed-form solutions in (\ref {Total_Rate_DL2}) and (\ref {Total_Rate_UL}) for both perfect and imperfect hardwares. In contrast to perfect hardware, the figure shows that the ESR ceiling phenomenon occurs for imperfect hardware which reveals the performance limits of hardware-constrained realistic networks in the high SNR regime. This figure also reveals that hardware imperfections have low impact at low SNRs, but are significant in the high SNR regime. Furthermore, it is observed that the proposed OPA increases the secrecy rate floor by approximately 1 bits/s/Hz and 0.9 bits/s/Hz for DL and UL scenarios, respectively compared to the EPA ($\lambda=0.5$).

In Fig. \ref {esr_4}, we examine the accuracy of the derived closed-form solutions for MA-S and MA-D by considering $N_\mathrm{s}=N_\mathrm{d}=4$. As can be seen, the numerical and the theoretical curves are in good agreement across all SNR regimes. Moreover, it is observed that by increasing the level of hardware imperfections from $k=0.05$ to $k=0.1$, the achievable secrecy rate is degraded approximately 1 bits/s/Hz in the high SNR regime.

Figs. \ref{sop_dl} and \ref{sop_ul} show the SOP as a function of the transmit SNR for LSMA based DL and UL scenarios, respectively, and for different target secrecy rates. The theoretical curves were plotted by the derived analytical expressions in (\ref{SOP_DL_rayleigh}) and (\ref{SOP_UL_rayleigh}) which are well-tight with the marker symbols generated by the Monte-Carlo simulations. As observed from these figures, there is only a negligible
performance loss caused by transceiver hardware imperfections in the low target secrecy rate of $R_t=0.25$ bits/s/Hz, but by increasing the target secrecy rate to $R_t=1$ bits/s/Hz or $R_t=2$ bits/s/Hz, substantial performance loss is revealed. Interestingly, for $R_t=2$ bits/s/Hz,
the network with imperfect hardware is always in outage and secure communications is unattainable-irrespective of the transmit SNR. This is exactly predicted by our analytical results in section V. The reason is that this target secrecy rate is more than the derived thresholds in (\ref{SOP_DL_rayleigh}) and (\ref{SOP_UL_rayleigh}), and as mentioned, the SOP of the system always equals one for $R_t$ more the thresholds. It can also be seen from the figures that despite the OPA technique that the SOP curves with imperfect hardware and with perfect hardware have the same slope (and thus, hardware imperfections lead to only an SNR offset which is unveiled as a curve shifting to the right), the SOP performance of the EPA technique approaches a non-zero saturation value in the high SNR regime for imperfect hardware. This observation reveals the secrecy performance advantage of the proposed OPA scheme compared with EPA.

Finally, we provide Figs. \ref{design_dl} and \ref{design_ul} to illustrate the insights for designing practical systems that were presented in Section VI. In the simulation, we assume that the total hardware imperfection over each node equals to 0.2, i.e., $k_\mathrm{R}^\mathrm{tot}=k_\mathrm{D}^\mathrm{tot}=0.2$. Based on Propositions 2 and 4, to maximize the secrecy rate, we should design the transmission and reception RF front ends at R such that $k_\mathrm{R}^t=k_\mathrm{R}^r=0.1$. For LSMA at S, based on Proposition 3, we obtain $k_\mathrm{D}^t=0.13$ and  $k_\mathrm{D}^r=0.07$ while for LSMA at D and based on Proposition 5, we should design the hardwares such that $k_\mathrm{D}^t=0.2$ and $k_\mathrm{D}^r=0$. By defining the hardware imperfection vector as $\mathrm{IV}=[k_\mathrm{R}^t, k_\mathrm{R}^r, k_\mathrm{D}^t, k_\mathrm{D}^r]$, we consider the following four different hardware design schemes:
\begin{itemize}
\item Design 1: R and D are designed randomly, for example IV$=[0.15, 0.05, 0.1, 0.1]$,
\item Design 2: R is designed optimally based on Propositions 2 and 4 while D is designed randomly; IV$=[0.1, 0.1, 0.1, 0.1]$,
\item Design 3: R is designed randomly while D is designed optimally; For LSMA at S, IV$=[0.15, 0.05, 0.13, 0.07]$ and for LSMA at D, IV$=[0.15, 0.05, 0.2, 0]$, and
\item Design 4: R and D are designed optimally; For LSMA at S, IV$=[0.1, 0.1, 0.13, 0.07]$ and for LSMA at D, IV$=[0.1, 0.1, 0.2, 0]$.
\end{itemize}
The results depict that the hardware design 4 which is based on Propositions 2-5 provides higher ESR performance compared to the case of random  hardware design (Design 1) and the cases of optimizing only one node (Designs 2 and 3). Furthermore, they show that the analysis presented in Section VI (which was based on high SNR analysis), can be utilized auspiciously at medium SNRs. In addition, as can be seen from these figures and mentioned before, different hardware designs have the ESR performance close together at low SNR regime, while the difference between the ESR performance of the designs is large at high SNR regime. Finally, we can understand from the figure that the proposed OPA together with Design 4 significantly outperforms the scenario of EPA with Design 4.

\section{Conclusion}
Physical radio-frequency (RF) transceivers are inseparable segments in both traditional and new emerging wireless networks. In the literature, very few works have considered the impact of hardware imperfections on security based transmissions and little is understood regarding this impact on untrusted relaying networks. In this paper, by taking hardware imperfections into consideration, we proposed an optimal power allocation (OPA) strategy to maximize the instantaneous secrecy rate of a cooperative wireless network comprised of a source, a destination and an untrusted amplify-and-forward (AF) relay. Based on our OPA solutions, new closed-form expressions were derived for the ergodic secrecy rate (ESR) and secrecy outage probability (SOP) with Rayleigh fading channels. The expressions effectively characterize the impact of hardware imperfections and manifest the existence of a secrecy rate ceiling that cannot be enhanced by increasing SNR or improving fading conditions. They also illustrate that hardware imperfections have low impact at low SNRs, but are significant in the high SNR regime. This issue reveals that hardware imperfections should be taken into account when developing high rate systems such as LTE-Advanced and 5G networks. To improve the secrecy performance of the network, we finally presented the hardware design approach. Numerical results depict that optimally distributing the hardware imperfections between the transmission and reception RF segments can further improve the secrecy performance.\\
\appendices

\vspace{-3mm}
\section{}\label{appA}
Let take the first-order derivative of $\phi^{\infty}$ on $k_\mathrm{R}^t$ using the chain rule in partial derivations as follows
\begin{align}\label{large_der_phi_tot}
&\frac{\partial \phi^{\infty}}{\partial k_\mathrm{R}^t}=\frac{\partial \phi^{\infty}}{\partial \theta_L }\Big(\frac{\partial \theta_L}{\partial \tau_1}\frac{\partial \tau_1 }{\partial k_\mathrm{R}^t}+\frac{\partial \theta_L}{\partial \tau_2}\frac{\partial \tau_2 }{\partial k_\mathrm{R}^t}+\frac{\partial \theta_L}{\partial \tau_3 }\frac{\partial \tau_3}{\partial k_\mathrm{R}^t}+\frac{\partial \theta_L}{\partial \xi_1 }\frac{\partial \xi_1}{\partial k_\mathrm{R}^t}\Big)\nonumber\\
&+\frac{\partial \phi^{\infty}}{\partial \tau_1}\frac{\partial \tau_1 }{\partial k_\mathrm{R}^t}+\frac{\partial \phi^{\infty}}{\partial \tau_2}\frac{\partial \tau_2 }{\partial k_\mathrm{R}^t}+\frac{\partial \phi^{\infty}}{\partial \tau_3 }\frac{\partial \tau_3}{\partial k_\mathrm{R}^t}+\frac{\partial \phi^{\infty}}{\partial \xi_1 }\frac{\partial \xi_1}{\partial k_\mathrm{R}^t},
\end{align}
where using (\ref {phi_infty}), we obtain
\begin{align}\label{der_phi_thetaL}
&\frac{\partial \phi^{\infty}}{\partial \theta_L}={\frac { \kappa_1 {{\theta_L}}^{2}+\kappa_2 \theta_L+ \kappa_3}{ ( \theta_L { \xi_1}+{
\tau_1} ) ^{2} ( { \tau_2}\theta_L+{
 \tau_3} ) ^{2}}},\\
 &\frac{\partial \theta_L}{\partial \tau_1}={\frac {\tau_3}{2\sqrt {{ {{\tau_2 \tau_3}
 \left( { \tau_1}-{\tau_3} \right) }}}}},\\
 &\frac{\partial \theta_L}{\partial \tau_2}=-\frac{\sqrt{\tau_3(\tau_1-\tau_3)}}{2\tau_2\sqrt{\tau_2}}-\frac{\tau_3(\xi_1-1)}{\tau_2^2},\\
 &\frac{\partial \theta_L}{\partial \tau_3}=\frac{\tau_1-2\tau_3}{2\sqrt{\tau_2\tau_3(\tau_1-\tau_3)}}+\frac{\xi_1-1}{\tau_2}-1,\\
 &\frac{\partial \theta_L}{\partial \xi_1}=\frac{\tau_3}{\tau_2},\\
 &\frac{\partial \phi^{\infty}}{\partial \tau_1}=\frac { { \theta_L}\left( {\tau_2}
\theta_L +{\tau_3}+{ \theta_L} \right) }{ (\theta_L{\xi_1}+{ \tau_1}) ^{2} \left(
{\tau_2}\theta_L+{ \tau_3} \right) }
,\\
 &\frac{\partial \phi^{\infty}}{\partial \tau_2}=-\frac{\theta_L^2(\xi_1 \theta_L+\tau_1-\theta_L)}{(\tau_2 \theta_L+\tau_3)^2(\xi_1 \theta_L+\tau_1)},\\
 &\frac{\partial \phi^{\infty}}{\partial \tau_3}=-\frac{\theta_L (\xi_1 \theta_L +\tau_1-\theta_L)}{(\tau_2 \theta_L+\tau_3)^2(\xi_1 \theta_L+\tau_1)},\\
 &\frac{\partial \phi^{\infty}}{\partial \xi_1}=\frac{\theta_L^2 (\tau_2 \theta_L +\tau_3+\theta_L)}{(\xi_1 \theta_L+\tau_1)^2(\tau_2 \theta_L+\tau_3)},\label{der_phi_xi1}
\end{align}
where $\kappa_1=-{ \tau_1}\tau_2(\tau_2+1 )+{\tau_3}{\xi_1} ( {\xi_1}-1 ) $, $\kappa_2=-2{\tau_1}{\tau_3} ( {\tau_2}-{\xi_1}+1 )$ and $\kappa_3={ \tau_1}{ \tau_3}( {\tau_1}-{
 \tau_3})$. By substituting  $k_\mathrm{R}^r=k_\mathrm{R}^\mathrm{tot}-k_\mathrm{R}^t$ into (\ref {phi_infty}), we obtain
\begin{align}\label{der_tau1_kRt}
&\frac{\partial \tau_1 }{\partial k_\mathrm{R}^t}=-2k_\mathrm{R}^{\mathrm{tot}}+2k_\mathrm{R}^t,\\
&\frac{\partial \tau_2 }{\partial k_\mathrm{R}^t}=-2{k_\mathrm{D}^r}^2(k_\mathrm{R}^\mathrm{tot}-k_\mathrm{R}^t)-2(k_\mathrm{R}^\mathrm{tot}-k_\mathrm{R}^t)
{k_\mathrm{R}^t}^2+2(k_\mathrm{R}^\mathrm{tot}-k_\mathrm{R}^t)^2k_\mathrm{R}^t+4k_\mathrm{R}^t-2k_\mathrm{R}^\mathrm{tot},\label{der_tau2_kRt}\\
&\frac{\partial \tau_3}{\partial k_\mathrm{R}^t}=-2{k_\mathrm{D}^r}^2(k_\mathrm{R}^\mathrm{tot}-k_\mathrm{R}^t)-2(k_\mathrm{R}^\mathrm{tot}-k_\mathrm{R}^t)
{k_\mathrm{R}^t}^2+2(k_\mathrm{R}^\mathrm{tot}-k_\mathrm{R}^t)^2k_\mathrm{R}^t+4k_\mathrm{R}^t-2k_\mathrm{R}^\mathrm{tot}+2 k_\mathrm{R}^t {k_\mathrm{D}^t}^2,\label{der_tau3_kRt}\\
&\frac{\partial \xi_1 }{\partial k_\mathrm{R}^t}=-2k_\mathrm{R}^\mathrm{tot}+2k_\mathrm{R}^t.\label{der_xi1_kRt}
\end{align}
Substituting (\ref {der_phi_thetaL})--(\ref {der_xi1_kRt}) into (\ref {large_der_phi_tot}) and after tedious manipulations yields
\begin{eqnarray}\label{pr_corr22}
\frac{\partial \phi^{\infty}}{\partial k_\mathrm{R}^t}=\frac{4 (1-{k_\mathrm{D}^r}^2) (k_\mathrm{R}^{\mathrm{tot}}-2 k_\mathrm{R}^t)}{\Big(4{k_\mathrm{R}^t}^2-4 k_\mathrm{R}^t k_\mathrm{R}^\mathrm{tot} + 2{k_\mathrm{R}^\mathrm{tot}}^2 +2{k_\mathrm{D}^r}^2+{k_\mathrm{D}^t}^2\Big)^2}.
\end{eqnarray}
Expression (\ref {pr_corr22}) shows that $\phi^{\infty}$ is a concave function of $k_\mathrm{R}^t$ in the feasible set and  $k_\mathrm{R}^t=\frac{k_\mathrm{R}^\mathrm{tot}}{2}$ is the single solution to $\frac{\partial \phi^{\infty}}{\partial k_\mathrm{R}^t}=0$.

\vspace{-1mm}
\section{}\label{appB}
\vspace{-1mm}Following the similar approach in {Proposition 2}, we should evaluate $\frac{\partial \phi^{\infty}}{\partial k_\mathrm{D}^t}$. Let substitute $k_\mathrm{D}^r=k_\mathrm{D}^{\mathrm{tot}}-k_\mathrm{D}^t$ into $\tau_1$, $\tau_2$, $\tau_3$ and then compute the following derivations \begin{align}\label{der_tau_kDt1}
&\frac{\partial \tau_1}{\partial k_\mathrm{D}^t}=1,~~\frac{\partial \tau_2}{\partial k_\mathrm{D}^t}=-2{k_\mathrm{R}^r}^2(k_\mathrm{D}^\mathrm{tot}-k_\mathrm{D}^t)-2k_\mathrm{D}^\mathrm{tot}+2k_\mathrm{D}^t,\\
&\frac{\partial \tau_3}{\partial k_\mathrm{D}^t}=-2{k_\mathrm{R}^r}^2(k_\mathrm{D}^\mathrm{tot}-k_\mathrm{D}^t)-2k_\mathrm{D}^\mathrm{tot}+2k_\mathrm{D}^t+
2k_\mathrm{D}^t({k_\mathrm{R}^t}^2+1)+2k_\mathrm{D}^t(k_\mathrm{D}^\mathrm{tot}-k_\mathrm{D}^t)^2-2{k_\mathrm{D}^t}^2(k_\mathrm{D}^\mathrm{tot}-k_\mathrm{D}^t).\label{der_tau_kDt3}
\end{align}
The expression $\frac{\partial \phi^{\infty}}{\partial k_\mathrm{D}^t}$ can be obtained similar to (\ref {large_der_phi_tot}) by changing ${ k_\mathrm{R}^t}$ to ${ k_\mathrm{D}^t}$. Then by substituting (\ref {der_phi_thetaL})--(\ref {der_phi_xi1}) and (\ref {der_tau_kDt1}), (\ref {der_tau_kDt3}) into $\frac{\partial \phi^{\infty}}{\partial k_\mathrm{D}^t}$, and after manipulations, we obtain
\begin{align}\label{proposition3}
\frac{\partial \phi^{\infty}}{\partial k_\mathrm{D}^t}=-\frac{2\Big(2k_\mathrm{R}^2 k_\mathrm{D}^t-2{k_\mathrm{D}^t}^2k_\mathrm{D}^\mathrm{tot}+2 k_\mathrm{D}^t{k_\mathrm{D}^\mathrm{tot}}^2+3k_\mathrm{D}^t-2k_\mathrm{D}^\mathrm{tot}\Big)}{\Big(2k_\mathrm{R}^2+3{k_\mathrm{D}^t}^2-4k_\mathrm{D}^tk_\mathrm{D}^\mathrm{tot}+2{ k_\mathrm{D}^\mathrm{tot}}^2\Big)^2}.
\end{align}
It is straightforward to see that (\ref {proposition3}) is a concave function of $ k_\mathrm{D}^t$ in the feasible set and the single solution to $\frac{\partial \phi^{\infty}}{\partial k_\mathrm{D}^t}=0$ is simply calculated.\\

\vspace{-3mm}
\section{}\label{appC}
We can write
\begin{eqnarray} \label{small_der_phi_tot}
\frac{\partial \phi^{\infty} }{\partial k_\mathrm{R}^r}=\frac{\partial \phi^{\infty}}{\partial \tau_2 } \frac{\partial \tau_2}{\partial k_\mathrm{R}^r}+\frac{\partial \phi^{\infty}}{\partial \xi_1 }\frac{\partial \xi_1}{\partial k_\mathrm{R}^r}.
\end{eqnarray}
Using (\ref {phi_infty}) yields
\begin{align}\label{small_der_phi_1}
&\frac{\partial \phi^{\infty}}{\partial \tau_2 }=-{\frac { \xi_1{\hspace{-.2mm}}\Big[\hspace{-.7mm}
(2\left( \tau_2\hspace{-.8mm}+\hspace{-1mm}2 \right) {\xi_1}\hspace{-1mm}-\hspace{-1mm}{\tau_2}^{2}
 ) {\hspace{-.4mm}}\sqrt {\xi_1(1\hspace{-1mm}+\hspace{-1mm}\tau_2)}\hspace{-1mm}+\hspace{-1mm}2\xi_1( \tau_2( \xi_1\hspace{-1mm}+\hspace{-1mm}1\hspace{-1mm}-\hspace{-1mm}\tau_2) \hspace{-1mm}+\hspace{-1mm}\xi_1\hspace{-1mm}+\hspace{-1mm}1) \Big] }{2 {\tau_2}^{2}
 \left( \sqrt {\xi_1}\sqrt {1\hspace{-1mm}+\hspace{-1mm}\tau_2}\hspace{-1mm}+1\hspace{-.5mm} \right) ^{2} \left( \xi_1\hspace{-1mm}+\hspace{-1mm}\sqrt {\xi_1
}\sqrt{1\hspace{-.7mm}+\hspace{-.7mm}\tau_2} \right) ^{2}}}, \\
&\frac{\partial \phi^{\infty}}{\partial \xi_1}={\frac { \left( 1+\tau_2 \right)  \Big[  \left( \tau_2+2{
\xi_1} \right) \sqrt {\xi_1(1+\tau_2)}+ 2\left( 1+\tau_2 \right) \xi_1 \Big]
}{2\tau_2 \left(
\sqrt {\xi_1}\sqrt {1+\tau_2}+1 \right) ^{2} \left( \xi_1+\sqrt {\xi_1}\sqrt {1+\tau_2} \right) ^{2} }}.
\end{align}
Considering $k_\mathrm{R}^t=k_\mathrm{R}^\mathrm{tot}-k_\mathrm{R}^r$, one can obtain
\begin{align}\label{small_der_phi_3}
&\frac{\partial \tau_2}{\partial k_\mathrm{R}^r}=4 {k_\mathrm{R}^r}^3-6{k_\mathrm{R}^r}^2{k_\mathrm{R}^\mathrm{tot}}+
(2{k_\mathrm{D}^r}^2+2{k_\mathrm{R}^\mathrm{tot}}^2+4){k_\mathrm{R}^r}-2{k_\mathrm{R}^\mathrm{tot}},\\
&\frac{\partial \xi_1}{\partial k_\mathrm{R}^r}=2{k_\mathrm{R}^r}.\label{small_der_phi_4}
\end{align}
By substituting (\ref {small_der_phi_1})--(\ref {small_der_phi_4}) into (\ref {small_der_phi_tot}) and solving $\frac{\partial \phi^{\infty} }{\partial k_\mathrm{R}^r}=0$ yields $k_\mathrm{R}^r=\frac{k_\mathrm{R}^\mathrm{tot}}{2}$.\\

~~~~~~~~~~~~~~~~~~~~~~~~~~~~~~~~~~~~~~~~~~~{\bf Acknowledgements}\\
The authors would like to thank Prof. Lajos Hanzo for helpful comments to improve the paper.

\end{document}